\documentclass[10pt]{article}

\usepackage{graphicx}              
\usepackage{amsmath}               
\usepackage{amssymb, amsbsy, latexsym} 
\usepackage{amsfonts}              
\usepackage[mathscr]{eucal}     
\usepackage{amsthm}                
\usepackage{multirow}    
\usepackage{txfonts}

\makeatletter
\@addtoreset{equation}{section}
\makeatother

\newtheorem{thm}{Theorem}[section]
\newtheorem{lem}[thm]{Lemma}

\newtheorem{cor}[thm]{Corollary}

\newcommand{\be}{\begin{eqnarray}}
\newcommand{\ee}{\end{eqnarray}}

\newcommand{\cd}{\mathcal D}

\newcommand{\cf}{\mathcal F}

\newcommand{\cv}{\mathcal V}

\newcommand{\cz}{\mathcal Z}

\renewcommand{\a}{\alpha}
\renewcommand{\b}{\beta}
\newcommand{\g}{\gamma}
\newcommand{\G}{\Gamma}

\newcommand{\eps}{\epsilon}

\renewcommand{\o}{\omega}

\newcommand{\rmd}{\mathrm d}

\newcommand{\startproof}{\setlength{\parindent}{0in}\textbf{Proof.} }
\newcommand{\finishproof}{\hfill $\blacksquare$ \\}

\def\tr{\mathrm{tr}}
\def\half{\frac{1}{2}}

\newcommand{\scrV}{\mathcal{V}}

\newcommand{\scrD}{\mathcal{D}}

\newcommand{\scrZ}{\mathcal{Z}}

\newcommand{\scrF}{\mathcal{F}}
\newcommand{\scrG}{\mathcal{G}}
\newcommand{\scrH}{\mathcal{H}}

\newcommand{\dual}{\,\,{}^\star\!}

\newcommand{\dummy}{\rule{0mm}{0mm}}

\newcommand{\topdec}[3]{\overset{\scriptscriptstyle #2}{\displaystyle
#1}\rule{0pt}{#3}}
\newcommand{\gdec}[1]{\topdec{#1}{(\gamma)}{8pt}}

\newcommand{\plform}{X}

\newcommand{\plconj}{B}
\newcommand{\conn}{\omega} 

\title{{\sf Canonical path integral measures for Holst and Plebanski
gravity.}\\
{\sf I. Reduced Phase Space Derivation}}

\author{
{\sf Jonathan Engle$^{1,2,4}$}\thanks{{\sf engle@theorie3.physik.uni-erlangen.de}},
{\sf Muxin Han$^{1,4}$}\thanks{{\sf
mhan@aei.mpg.de}},
{\sf Thomas Thiemann$^{1,3,4}$}\thanks{{\sf
thiemann@theorie3.physik.uni-erlangen.de, thiemann@aei.mpg.de,
tthiemann@perimeterinstitute.ca}}\\
\\
{\sf $^1$ MPI f. Gravitationsphysik, Albert-Einstein-Institut,} \\
           {\sf Am M\"uhlenberg 1, 14476 Potsdam, Germany}\\
\\
{\sf $^2$ Centre de Physique Th\'eorique\footnote{Unit\'e Mixte
de Recherche (UMR 6207) du CNRS et des Universit\'es Aix-Marseille
I, Aix-Marseille II, et du Sud Toulon-Var; laboratoire afili\'e
\`a la FRUMAM (FR 2291)}} \\
{\sf Campus de Luminy, Case 907, 13288 Marseille, France}\\
\\
{\sf $^3$ Perimeter Institute for Theoretical Physics,} \\
{\sf 31 Caroline Street N, Waterloo, ON N2L 2Y5, Canada}\\
\\
{\sf $^4$ Institut f. Theoretische Physik III, Universit\"{a}t
Erlangen-N\"{u}rnberg} \\
{\sf Staudtstra{\ss}e 7, 91058 Erlangen, Germany}
}
\date{}

\begin{document}

\maketitle

\begin{abstract}
{\sf
An important aspect in defining a path integral quantum theory is the
determination of the correct measure.  For interacting theories and
theories with constraints, this is non-trivial, and is normally not the
heuristic "Lebesgue measure" usually used. There have been many
determinations of a measure for gravity in the literature, but none for
the Palatini or Holst formulations of gravity. Furthermore, the
relations between different resulting measures for different
formulations of gravity are usually not discussed.

In this paper we use the reduced phase technique in order to derive the
path-integral measure for the Palatini and Holst formulation of gravity,
which is different from the Lebesgue measure up to local measure factors
which depend on the spacetime volume element and spatial volume element.

From this path integral for the Holst formulation of GR we can
also give a new derivation of the Plebanski path integral and discover
a discrepancy with the result due to Buffenoir, Henneaux, Noui
and Roche (BHNR) whose origin we resolve. This paper is the first in a
series that aims at better understanding the relation between canonical
LQG and the spin foam approach.
}
\end{abstract}

\newpage

\tableofcontents

\newpage

\section{Introduction}

Richard Feynman, in the course of his doctoral work, developed
the path integral formulation of quantum mechanics as an alternative,
space-time covariant description of quantum mechanics, which is
nevertheless equivalent to the canonical approach \cite{feynman}.  It is
thus not surprising that the path integral formulation
has been of interest in the quantization of general relativity, a theory
where space-time
covariance plays a key role.
However, once one departs from the regime of free, unconstrained
systems, the equivalence of the path integral approach and canonical
approach becomes
more subtle than originally described by Feynman in \cite{feynman}.   In
particular, in Feynman's
original argument, the integration measure for the configuration path
integral is a formal Lebesgue measure; in the interacting case, however,
in order to have
equivalence with the canonical theory, one cannot use the naive Lebesgue
measure
in the path integral, but must use a measure derived from the Liouville
measure
on the phase space \cite{HT}.

Such a measure has yet to be incorporated into spin-foam models, which
can be thought
of as a path-integral version of loop quantum gravity (LQG)
\cite{sfrevs, rr97}.
Loop quantum gravity is an attempt to make a mathematically
rigorous quantization of general relativity that preserves
background independence --- for reviews, see \cite{thiemannrev, alrev,
smolinrev} and for books see \cite{rovellibook, thiemannbook}.
Spin-foams intend to be a path integral formulation for loop quantum gravity, directly motivated from the ideas
of Feynman appropriately adapted to reparametrization-invariant theories \cite{rr97, reisenberger}.
Only the kinematical structure of LQG is used in motivating the spin-foam framework. The
dynamics one tries to encode in the
amplitude factors appearing in the path integral which is being replaced
by a sum in a regularisation step which depends on a triangulation of
the spacetime manifold. Eventually one has to take a
weighted average over these (generalised) triangulations for which
the proposal at present is to use methods from group field theory
\cite{sfrevs}. The current spin foam approach is independent from the dynamical theory of canonical LQG
\cite{QSD} because the dynamics of canonical LQG is rather complicated. It instead uses
an apparently much simpler starting point:
Namely, in the Plebanski formulation \cite{plebanski}, GR can be
considered as a
constrained BF theory, and treating the so called simplicity
constraints as a perturbation of BF theory, one can make use of the
powerful toolbox that comes with topological QFT's \cite{bf}. It is
an unanswered question, however, and one of the most active research
topics momentarily\footnote{Here we are referring the spin-foam model for 4-dimensional gravity, while for 3-dimensional gravity the consistency is discussed in e.g. \cite{perez}.}, how canonical LQG and spin foams fit together.
It is one the aims of this paper to make a contribution towards
answering this question.

In LQG one is compelled to introduce a 1-parameter quantization
ambiguity --- the so-called Immirzi parameter \cite{barbero,immirzi}.
This enters the
action through a necessary extra `topological' term added to the
Palatini action; the full
action is termed the \textit{Holst} action \cite{holst}. To properly
incorporate
the Immirzi parameter into spin-foams,
one should in fact not start from the usual Plebanski formulation but
rather an analogous
generalization, in which an analogous topological term is added to the
action, leading to
what we call the Plebanski-Holst formulation of gravity
\cite{CMPR,LO,spinfoam2}.

In \cite{companion} we have shown (and partly
reviewed) for a rather general
theory that different canonical quantisation techniques for gauge
theories,
specifically Dirac's operator constraint method, the Master Constraint
method and the reduced phase space method all lead to the same path
integral. A prominent role in establishing this equivalence is played
by what is called ``the choice of gauge fixing'' (from the reduced
phase space point of view) or, equivalently, the choice of clocks (from
the gauge invariant i.e. relational point of view \cite{observable}).
After a long analysis, it transpires that the common basis for the path
integral measure, no matter from which starting point it is derived, is
the Liouville measure on the reduced phase, which can be defined via
gauge fixing of the first class constraints. This measure can be
extended to the full phase space and one shows that
the dependence on the gauge fixing disappears when one integrates
gauge invariant functions\footnote{However, the dependence on the gauge
fixing is secretly there, {\it in a gauge invariant form}, since choices of
algebras of Dirac observables (i.e. gauge invariant functions)
are in one to one correspondence with choices of gauge fixing. The choice of
such an algebra is the zeroth step in a canonical quantisation scheme
and determines everything else such as the representation theory,
see \cite{companion} for a comprehensive discussion.}.
From this point of view, that is,
the equivalence between path-integral formulation and the canonical
theory, it is obvious that formal path-integrals derived from the
various formulations of gravity should all be equivalent, because all of
them have the same reduced phase space --- that of general relativity.

We thus apply the general reduced phase space framework
to the Holst action as the starting point for
deriving a formal path integral for \textit{both} the Holst action and
the Plebanski-Holst action. It turns out that the resulting
path-integral for either the Holst action or the Plebanski-Holst action
is not the naive Lebesgue measure integral of the exponentiated
action.
There are extra measure factors of spacetime volume element $\cv$ and
spatial volume element $V_s$. The presence of a spatial volume element
is
especially surprising because it breaks the manifest spacetime
covariance of the
path-integral when we are off shell. The origin of this
lack of
covariance is in the mixture of dynamics and gauge invariance
inherent to generally covariant systems with propagating degrees of freedom
and it is well known that
the gauge symmetries generated by the constraints only coincide on shell
with spacetime diffeomorphism invariance. The quantum theory chooses to
preserve the gauge symmetries generated by the constraints rather than
spacetime diffeomorphism invariance when we take quantum corrections
into account (go off shell).

This kind of extra measure factor (so called local measure) has appeared
and been discussed in the literature since 1960s (see for instance
\cite{FV,GT}) in the formalism of geometrodynamics and its {\it
background-dependent} quantizations (stationary phase approximation).
The outcome from the earlier investigations appears to be that in
background-dependent, perturbative quantizations, these measure factors
of $\cv$ and
$V_s$ only contribute to the divergent part of the higher loop-order
amplitudes. Thus their meanings essentially depend on the regularization
scheme used.
One can of course try to choose certain regularization
schemes such that, either the local measure factors never contribute to
the
transition amplitude, or that their effect is canceled by the divergence
from the
action \cite{FV,GT}. However, the power of renormalisation and the very
reason we trust it is that its predictions are independent of the
regularisation technique chosen. Therefore the status of these measure
factors is very much unsettled, especially for non perturbative
quantisation techniques. We here take the point of view that the measure
factors should be taken seriously because they take the off shell
symmetry generated by the constraints properly into account. In which
sense this so called Bergmann -- Komar ``group'' \cite{BK} is preserved
in the
path integral is the subject of the research conducted in \cite{muxin4}.
In this article we confine ourselves to a brief discussion.

In the formalism of connection-dynamics, which is a preparation of
background-independent quantization, a similar local measure factor also
appears. It was first pointed out in \cite{bhnr}, whose path-integral
will be shown to be equivalent to our present formulation up to a
discrepancy whose origin we resolve.
When we perform {\it background-independent} quantization as in
spin-foam models, therefore the local measure factor should not be
simply ignored, because the regularization arguments in
background-dependent quantization have no obvious bearing in the
background-independent context anymore. For example, spin-foam models
are defined on a triangulation of the spacetime manifold with finite
number of vertices, where at each vertex the value of local measure is
finite, and the action also does not show any divergence.

However, so far none of the existing spin-foam models implements this
non-trivial local measure factor in the quantization
\footnote{The ambiguities of the path integral measure in spin-foam models have been discussed in the literatures. In the context of spin-foam models, this issue of path integral measure can be translated into an ambiguity of the gluing amplitudes between 4-cells \cite{measre}, while the quantum effects are discussed in \cite{measure1}. And the relevance of the measure factor on the diffeomorphism symmetries of the spinfoam amplitudes is discussed in \cite{measure2}. However in the present work we are concerning the measure factor which helps to make a connection with the canonical quantization, while the standard spin-foam approach doesn't rely on the canonical framework and 3+1 splitting of the spacetime manifold.}.
The quantum effect implied by this measure factor has not been analyzed in the context of
spin-foam models. But without it there is no chance to link spin foams
with canonical LQG which at present is the only method we have in order
to derive a path integral formulation of LQG from first principles.
In ongoing work
\cite{muxin2} we analyse the non-trivial effects
caused by this measure factor in the context of spin-foam models, and
try to give spin-foam amplitudes an unambiguous canonical
interpretation by establishing a link between path-integral formulation
and canonical quantization. In this article we also make a few comments
on this.\\
\\
The paper is organized as follows: \\
\\

In section \ref{holst_sect}, after defining the reduced phase
space path integral for a general theory, we begin with
the Hamiltonian framework
arising from the $SO(\eta)$ Holst action\footnote{Our discussions apply
to both Euclidean and Lorentzian signatures.} \cite{barros} (see also
\cite{alexandrov}). We then derive the path-integral formula for the
Holst
action in terms of spacetime field variables, i.e. the $so(\eta)$
connection $\o_\mu^{IJ}$ and the co-tetrad $e_\mu^I$.

In section \ref{pleb_sect}, starting from the Holst phase space path
integral, we construct a path-integral formula for the Plebanski-Holst
action by adding some extra fields and extra constraints.

In section \ref{consisBHNR}, we discuss the consistency with the
calculations in \cite{bhnr}.

Finally, we summarize and conclude with an outlook to future research.

\section{The path-integral measure for the Holst action}
\label{holst_sect}

\subsection{Reduced Phase space path integral}
\label{holst_phase}

To cut a long story short (see e.g. \cite{HT,companion}) the central
ingredient for most applications of the path integral is the generating
functional
\be \label{1}
\scrZ[j]:=\int\;\scrD q\;\scrD p\;
|\det(\{F,\xi\})|\;
\sqrt{\det(\{S,S\})}\; \delta[S]\;\delta[F]\;\delta[\xi]\;
\exp(i\int\;dt\;[p_a\dot{q}^a+j_a q^a])
\ee
Here $(q^a,p_a)$ denotes any instantaneous Darboux coordinates on phase
space, $S$
denotes
the collection of all second class constraints $S_\Sigma$, $F$
the
collection of all first class constraints $F_\mu$, $\xi$ any choice of
gauge
fixing conditions $\xi^\mu$, and $j$ is a current which allows us to
perform
functional derivations at $j=0$ in order to define any object of
physical interest. For instance the rigging kernel between initial and
final
kinematical states $\psi_i(q),\;\psi_f(q)$ results by generating these
two functions\footnote{Provided they are analytic. In case they are not,
they are analytic functions times a reference vector $\Omega_0$ in which
case the reference vector must be included in (\ref{1}). See \cite{companion}
for details.} through
functional derivation at $t=\pm \infty$. In addition, as usual $\scrD
q=\prod_{t\in \mathbb{R},a} dq^a(t)$ and $\delta[F]=\prod_{t\in
\mathbb{R},\mu} \delta(F_\mu(t))$, and likewise for $\scrD p$ and $\delta[S]$. We will
often write $|D_1|=[\det({F,\xi})]^2,\;|D_2|=\det(\{S,S\})$. We
will also drop the exponential of the current in what follows since, as long as it is a current 
multiplied into the tetrad variables,
it does not affect any of our manipulations --- hence we will mostly deal
with
the partition function $\scrZ=\scrZ[0]$. Since what one is really
interested in is $\scrZ[j]/\scrZ$ we can drop overall constant factors
from all subsequent formulas.

Applied to our situation,
we restrict ourself to the case of pure gravity defined by the Holst
action. We follow the notation employed in \cite{holst,barros}.
Note that for the simplicity of the formulae, we skip ``$\prod_{x\in
M}$" in almost all following path-integrals, where $M$ is the spacetime
manifold.
Moreover, we will assume
that all the gauge fixing conditions $\xi_\a$ are functions independent
of the connections $\o_a^{IJ}$ i.e. they are the functions of tetrad
only. This assumption will simplify the following discussion. Then
\be\label{holstphase}
\scrZ=\int \scrD\conn_a^{IJ} \scrD\pi_{IJ}^a\ \delta(C^{ab})\
\delta(D^{ab})\ \sqrt{|D_2|}\ \delta(G^{IJ})\ \delta(H_a)\ \delta(H)\
\sqrt{|D_1|}\prod_{\a}\delta(\xi_\a)\ \exp i\int\rmd t \rmd^3x\
\gdec{\pi}_{IJ}^a\dot{\conn}_a^{IJ}\label{pi}
\ee
where $\gdec{\pi}^a_{IJ} := (\pi - \frac{1}{\gamma} \dual \pi)^a_{IJ}$,
and the
expressions of the constraints $G^{IJ}$, $H_a$, $H$, $C^{ab}$, and
$D^{ab}$ are given by
\cite{barros}
\be
G_{IJ}&=&
D_a \gdec{\pi}^a_{IJ} :=
\partial_a \gdec{\pi}^a_{IJ}+\conn_{aI}^{\ \ K}\gdec{\pi}^a_{JK}
-\conn_{aJ}^{\ \ K} \gdec{\pi}^a_{IK}\nonumber\\
H_a&=&\frac{1}{2}F_{ab}^{IJ}[\conn]\ \gdec{\pi}^b_{IJ}\nonumber\\
H&=&\frac{1}{4\sqrt{\det q}}(F-\frac{1}{\g}*F)_{ab}^{IJ}[\conn]\
\pi^a_{IK}\ \pi^b_{JL}\ \eta^{KL}\nonumber\\
C^{ab}&=& \eps^{IJKL}\pi^a_{IJ}\pi^b_{KL}\nonumber\\
\label{holstconstr}
D^{ab}&=&\frac{1}{2\sqrt{\det
q}}*\pi^{c}_{IJ}(\pi^{aIK}D_c\pi^{bJL}+\pi^{bIK}D_c\pi^{aJL})\eta_{KL}
\ee
where $D^{ab}$ is the secondary constraint with
$\{H(x),C^{ab}(x')\}=D^{ab}(x)\delta(x,x')$. Note that the definition of
$H$ and $D^{ab}$ is slightly different from \cite{barros}, up to a
factor of
$1/(2\sqrt{\det q})$.
In rewriting the kinematical Liouville measure
$\scrD \conn^{IJ}_a \scrD \gdec{\pi}^a_{IJ}$ as
$\scrD \conn^{IJ}_a \scrD \pi^a_{IJ}$, an overall constant Jacobian
factor has also been dropped.

In Eq.(\ref{pi}) $D_2$ is the determinant of the Dirac matrix
\be
\begin{pmatrix}
      &\{C^{ab}(x),C^{cd}(x')\}&,\  \{C^{ab}(x),D^{cd}(x')\}\ \  \\
      &\{D^{ab}(x),C^{cd}(x')\}&,\  \{D^{ab}(x),D^{cd}(x')\}\ \
\end{pmatrix}
=
\begin{pmatrix}
      &0&,\  \{C^{ab}(x),D^{cd}(x')\}\ \  \\
      &\{D^{ab}(x),C^{cd}(x')\}&,\  \{D^{ab}(x),D^{cd}(x')\}\ \
\end{pmatrix}
\ee
Therefore $|D_2|=[\det G]^2$ where $G$ is the matrix
\be
G^{ab,cd}(x,x')=\{C^{ab}(x),D^{cd}(x')\}\approx (\det
q)^{3/2}\left[q^{ab}q^{cd}-
\half q^{ac}q^{bd}- \half q^{cb}q^{ad}\right]\ \delta^3(x,x').\label{G}
\ee
By the symmetry of $q^{ab}$, there exists an orthogonal
matrix\footnote{Here
and later on in the paper,
when we say a matrix is `orthogonal', even if it has spatial-manifold
indicies,
we mean orthogonal in the standard matrix sense -- i.e., `orthogonal'
with respect to $\delta_{ab}$, and not with respect to some covariantly
determined metric.}
$M^a{}_b$ such that
$M^a{}_c M^b{}_d q^{cd} = \lambda^a \delta^{ab}$ for some
$\{\lambda^a\}$, so that
\begin{equation}
M^a{}_e M^b{}_f M^c{}_g M^d{}_h G^{ef,gh}
= (\det q)^{3/2}\left[\lambda^a \lambda^c \delta^{ab} \delta^{cd}
- \half \lambda^a \lambda^b \delta^{ac} \delta^{bd}
- \half \lambda^c \lambda^a \delta^{cb} \delta^{ad}\right] .
\end{equation}
Let $\hat{G}^{ab,cd}:= \lambda^a \lambda^c \delta^{ab} \delta^{cd}
- \half \lambda^a \lambda^b \delta^{ac} \delta^{bd}
- \half \lambda^c \lambda^a \delta^{cb} \delta^{ad}$ denote the portion
in square brackets.
Each of the rows $(12), (13), (23)$ in $\hat{G}$
has exactly one non-zero matrix element. Reducing
$\det \hat{G}$ by minors along these 3
rows,
\begin{eqnarray}
\nonumber
\det \hat{G} &=& \hat{G}^{12,12} \hat{G}^{23,23} \hat{G}^{13,13} \det R
= - 2^{-9} \left(\lambda^1 \lambda^2 \lambda^3\right)^2 \det R
\end{eqnarray}
where $R^{ab} = \hat{G}^{aa,bb}$ is the reduced 3 by 3 matrix.
$\det R$ has only 2 non-zero terms; evaluating it and substituting in
the result gives
\begin{equation}
\det \hat{G} =  \frac{-1}{4} \left(\lambda^1 \lambda^2
\lambda^3\right)^4 = \frac{-1}{4}
(\det q^{ab})^4 = \frac{-1}{4} (\det q)^{-4},
\end{equation}
so that
\begin{equation}
\det G = \frac{-1}{4}(\det q)^{\frac{3}{2}\times 6}(\det q)^{-4} =
\frac{-1}{4}(\det q)^5.
\end{equation}
Thus, up to an overall factor,
\begin{equation}
\sqrt{|D_2|}=(\det q)^{5} =: V_s^{10}.
\end{equation}

Next we express the delta functions $\delta(H)$ and $\delta(D^{ab})$ in
Eq.(\ref{pi})
as integrals of exponentials,
\be
\scrZ&=&\int \scrD\conn_a^{IJ} \scrD\pi_{IJ}^a \scrD N \scrD d_{ab}\
V_s^{10}\ \delta(G^{IJ})\ \delta(H_a)\ \delta(C^{ab})\
\sqrt{|D_1|}\prod_{\a}\delta(\xi_\a)\ \exp i\int\rmd t \rmd^3x\
\left[\gdec{\pi}_{IJ}^a\dot{\conn}_a^{IJ}-NH+d_{ab}D^{ab}\right].\label{pi1}
\ee
Then we follow the strategy used in \cite{bhnr} to eliminate the
secondary second class constraint $D^{ab}$ in the path-integral. We
consider a change of variables which is also a canonical transformation
generated by the functional
\be
F := -\int\rmd^3x\ d_{ab}C^{ab}/N .
\ee
The integral measure is the Liouville measure on the phase space and
thus is invariant under canonical transformation. $\sqrt{|D_2|}$,
$G_{IJ}$, $C^{ab}$, and $\xi_{\a}$ are invariant because
they strongly Poisson commute with $C^{ab}$ (here we use the assumption
that the gauge fixing conditions $\xi_\a$ only depend on $\pi^a_{IJ}$),
and $H_a$ is invariant because it weakly
Poisson commutes with $C^{ab}$. $\sqrt{|D_1|}$ is not invariant under the canonical transformation, but the correction depends linearly on $d_{ab}$. Thus the correction will vanish under the Gauss integral over $d_{ab}$, which is performed later \footnote{One might be worried at first about the absolute value signs around this determinant in the path integral. However, as this Faddeev-Popov determinant should never be zero, it should never change sign, so that in fact the absolute value sign can just be removed.}. The change of kinetic term
$\delta\int\rmd t \rmd^3x\
\gdec{\pi}_{IJ}^a\partial_t{\conn}_a^{IJ}(x,t)$ is proportional to
$\int\rmd t \rmd^3x\ C^{ab}\partial_t(d_{ab}/N)$ which also vanishes by
the delta functions
$\delta(C^{ab})$ in front of the exponential. So $H$ and $D^{ab}$ are
the only terms that change in the canonical transformation generated by
$F$. Moreover because $\{H(x),C^{ab}(x')\}=D^{ab}(x)\delta(x,x')$ and
$\{C^{ab}(x),D^{cd}(x')\}=G^{ab,cd}(x,x')$ we can obtain explicitly the
transformation behavior of $H(N)$ and $D^{cd}(d_{cd})$
\be
\tilde{H}(N)&\equiv&\sum_{n=0}^\infty\frac{1}{n!}\{F,H(N)\}_{(n)}\nonumber\\
&=&\int\rmd^3x\ N(x)H(x)-\int\rmd^3x\int\rmd^3y\
\frac{N(x)}{N(y)}d_{ab}(y)\{C^{ab}(y),H(x)\}\nonumber\\
&+&\frac{1}{2}\int\rmd^3x\int\rmd^3y\int\rmd^3z\frac{N(x)}{N(y)N(z)}d_{ab}(y)d_{cd}(z)\{C^{ab}(y),\{C^{cd}(z),H(x)\}\}\nonumber\\
&=&\int\rmd^3x\ N(x)H(x)+\int\rmd^3x\int\rmd^3y\
\frac{N(x)}{N(y)}d_{ab}(y)D^{ab}(x)\delta(x,y)\nonumber\\
&-&\frac{1}{2}\int\rmd^3x\int\rmd^3y\int\rmd^3z\frac{N(x)}{N(y)N(z)}d_{ab}(y)d_{cd}(z)G^{ab,cd}(y,z)\delta(x,z)\nonumber\\
&=&\int\rmd^3x\ N(x)H(x)+\int\rmd^3x\
d_{ab}(x)D^{ab}(x)-\frac{1}{2}\int\rmd^3y\int\rmd^3z\frac{1}{N(y)}d_{ab}(y)d_{cd}(z)G^{ab,cd}(y,z)\\
\tilde{D}^{cd}(d_{cd})&\equiv&\sum_{n=0}^\infty\frac{1}{n!}\{F,D^{cd}(d_{cd})\}_{(n)}\nonumber\\
&=&\int\rmd^3x\ d_{cd}(x)D^{cd}(x)-\int\rmd^3x\int\rmd^3y\
\frac{1}{N(y)}d_{cd}(x)d_{ab}(y)\{C^{ab}(y),D^{cd}(x)\}\nonumber\\
&=&\ \int\rmd^3x\ d_{cd}(x)D^{cd}(x)-\int\rmd^3x\int\rmd^3y\
\frac{1}{N(y)}d_{cd}(x)d_{ab}(y)G^{ab,cd}(x,y)
\ee
here the series terminated because of $\{C^{ab}(x),
G^{cd,ef}(x',x'')\}=0$. Since $G^{ab,cd}(x,y)$ is proportional to
$\delta(x,y)$ we have
\be
\scrZ&=&\int \scrD\conn_a^{IJ} \scrD \pi_{IJ}^a \scrD N \scrD d_{ab}\
V_s^{10}\ \delta(G^{IJ})\ \delta(H_a)\ \delta(C^{ab})\
\sqrt{|D_1|}\prod_{\a}\delta(\xi_\a)\ \exp i\int\rmd t \rmd^3x\
\left[\gdec{\pi}_{IJ}^a\dot{\conn}_a^{IJ}-NH-\frac{1}{2}d_{ab}d_{cd}G^{ab,cd}/N\right]\nonumber\\
&=&\int \scrD\conn_a^{IJ} \scrD \conn_t^{IJ} \scrD\pi_{IJ}^a \scrD N^a
\scrD N \ \delta(C^{ab})\
\frac{V_s^{10}}{\sqrt{\left|\det (G/N)\right|}}\
\sqrt{|D_1|}\prod_{\a}\delta(\xi_\a)\ \exp i\int\rmd t \rmd^3x\
\left[\gdec{\pi}_{IJ}^a\dot{\conn}_a^{IJ}-\conn_t^{IJ}G_{IJ}-N^a
H_a-NH\right]\nonumber\\
\label{holstcanint}
&=&\int \scrD \conn_a^{IJ} \scrD\conn_t^{IJ} \scrD\pi_{IJ}^a \scrD N^a
\scrD N\ \delta(C^{ab})\ N^3
V_s^{5} \ \sqrt{|D_1|}\prod_{\a}\delta(\xi_\a)\ \exp i\int\rmd t
\rmd^3x\
\left[\gdec{\pi}_{IJ}^a\dot{\conn}_a^{IJ}-\conn_t^{IJ}G_{IJ}-N^aH_a-NH\right].\label{pi2}
\ee
This is the canonical phase space path integral for the Holst action,
with secondary
constraints removed as in \cite{hs}.  The Palatini
case is recovered by setting $\gamma = \infty$ while holding $G$
constant.

\subsection{Configuration path integral in terms of spacetime
$so(\eta)$-connection and tetrad}
\label{config_holst}

It is too difficult in concretely performing the integrations in
Eq.(\ref{pi2}) to compute transition amplitudes. However if we
transform the Eq.(\ref{pi2}) to be an integral of the Lagrangian Holst
action in terms of original configuration variables, i.e. the spacetime
connection field $\omega_\mu^{IJ}$ and tetrad field $e_\mu^I$, the integral
will become easier to handle. To rewrite the the canonical path
integral as a configuration path-integral for the Holst action, one
proceeds in two steps: (1.) Replace the canonical variables and Lagrange
multipliers with space-time variables and the simplicity constraint (2.)
Integrate out the simplicity constraint.

\subsubsection{Basic relations between variables}
\label{holst_basic}

In this section we give the definitions of the new coordinates
in terms of the old coordinates.
These definitions will be motivated and explained, and the bijectivity
of the coordinate transformation demonstrated, in the subsequent
section.

When the simplicity constraint is imposed,
\begin{equation}
C^{ab} = \eps^{IJKL}\pi^a_{IJ}\pi^b_{KL} \approx 0
\end{equation}
$\pi^a_{IJ}$ takes one of the five forms\footnote{To
see that the four sectors $(I \pm)$ and $(II \pm)$ are disjoint, define
$\pi^a_i := \half \pi^a_{0i}$ and $\tilde{\pi}^a_i:= \frac{1}{4}
\epsilon_i{}^{jk} \pi^a_{jk}$.
Then one has
\begin{eqnarray*}
(I+) &\Rightarrow& \det \pi^a_i = 0\text{ and }(\det
\tilde{\pi}^a_i)(\det e^i_a) > 0, \\
(I-) &\Rightarrow& \det \pi^a_i = 0\text{ and }(\det
\tilde{\pi}^a_i)(\det e^i_a) < 0, \\
(II+) &\Rightarrow& \det \tilde{\pi}^a_i = 0\text{ and }(\det
\pi^a_i)(\det e^i_a) > 0, \\
(II-) &\Rightarrow& \det \tilde{\pi}^a_i = 0\text{ and }(\det
\pi^a_i)(\det e^i_a) < 0 .
\end{eqnarray*}
}
\begin{eqnarray}
\nonumber
(I \pm) && \pi^a_{IJ} = \pm \eps^{abc} e_b^I e_c^J \\
\label{simp}
(II \pm) && \pi^a_{IJ} = \pm \half \eps^{abc} e_b^K e_c^L\eps_{IJKL}\\
(Deg) && \pi^a_{IJ} = 0.
\end{eqnarray}
Note that the appearance of the degenerated sector shows that the
Hamiltonian constrained system derived from the Holst action is not
regular, i.e. the rank of the Dirac matrix
\be
\begin{pmatrix}
      &\{C^{ab}(x),C^{cd}(x')\}&,\  \{C^{ab}(x),D^{cd}(x')\}\ \  \\
      &\{D^{ab}(x),C^{cd}(x')\}&,\  \{D^{ab}(x),D^{cd}(x')\}\ \
\end{pmatrix}
\ee
is not a constant on the whole phase space. We have to remove the
degenerated sector in order to carry out the derivations in the reduced
phase space. Therefore all the derivations in the last subsection hold
only if the degenerate sector is removed.
Now we restrict ourself in sector $(II+)$, and in addition stipulate
$\det e^i_a>0$, removing the sign ambiguity in the definition of
$e^I_a$. The derivations for other sectors can be carried out in the
same way. With the restriction to $(II+)$, the above relation can be
inverted as
\begin{equation}
e^I_a = \frac{1}{4\sqrt{2}}
\left|\det \pi^b_{oj} \right|^{-\half}
\epsilon^{IJKL} \epsilon_{abc} \pi^b_{0J} \pi^c_{KL} .
\end{equation}

This equation can then be used to define $e^I_a$ off-shell with respect
to the simplicity
constraint. One might ask whether $e^I_a$ so defined, along with
$C^{ab}$,
form good coordinates on $\pi^a_{IJ}$.   In fact, with the restrictions
just stipulated,
we will show $\pi^a_{IJ} \mapsto (e_a^I, C^{ab})$ is bijective
in the next subsection.

Lastly, we equip the internal space with
a time orientation, and define $n^I$ as the unique
internal future-pointing unit vector satisfying $n^I \pi^a_{IJ} = 0$.
Then one defines
\be
e_t^I := Nn^I+N^a e_a^I.
\ee
When the simplicity constraint is satisfied, $e_t^I$ is equal to
the $t$ component of the physical space-time tetrad, so that the above
definition is indeed an extension of the usual $e^I_t$.

\subsubsection{Proof of bijectivity}
\label{holst_bij}

For the purpose of making apparent the bijectivity of the coordinate
transformation,
and to aid in later calculations, define
\begin{eqnarray}
\pi^a_i &:=& \half \pi^a_{0i} \\
\tilde{\pi}^a_i &:=& \frac{1}{4} \epsilon_i{}^{jk} \pi^a_{jk} .
\end{eqnarray}
In terms of these, the `triad' $e^I_a$ defined in the last subsection
can be alternatively introduced via
\begin{eqnarray*}
\label{fdef}
1.& f^a_i :=& \left|\det \pi^b_j\right|^{-\half}
\pi^{a}_i, \qquad e^i_a = (f^a_i)^{-1} \\
2.& e^0_b :=& \half \epsilon_{abc} f^b_i \tilde{\pi}^c_i.
\end{eqnarray*}
Note that here $e^i_a$ denotes simply the
$I=1,2,3$ components of $e^I_a$, \textit{not} the co-triad.\footnote{One
can see that $e^i_a$ may not be taken as the co-triad from the
following. For $v^a, w^b$ tangent to the spatial slice $M$,
\begin{displaymath}
e^i_a e_{bi} v^a w^b = e^I_a e_{bI} v^a w^b + e^0_a e_{b0} v^a w^b
= g_{ab} v^a w^b + \sigma^2 (e^0_a v^a)(e^0_b w^b)
\equiv q_{ab} v^a w^b + s (e^0_a v^a)(e^0_b w^b)
\end{displaymath}
however, $e^0_a$ is in general arbitrary, so that the second term
on the right hand side is in general non-zero, whence in general
\begin{displaymath}
e^i_a e_{ai} \neq q_{ab} .
\end{displaymath}
}

The map $f^a_i \mapsto \pi^a_i$ is manifestly bijective.
The definition of $e^0_b$
uses precisely the information contained in
the anti-symmetric part of $f^a_i \tilde{\pi}^{bi}$,
whence the remaining information in
$\pi^a_{IJ}$ is exactly the symmetric part of
$f^a_i \tilde{\pi}^{bi}$:
\begin{equation}
S^{ab} := f^{(a}_i \tilde{\pi}^{b)i}
\end{equation}
In terms of this, the simplicity constraint is given by
\begin{equation}
\label{simpS}
C^{ab} = -2 \pi^{(a}_j \tilde{\pi}^{b)j} = -2 (\det e^k_c) S^{ab}.
\end{equation}
From this one sees that
$\pi^a_{IJ} \mapsto (e^I_a, C^{ab})$ is bijective.

\subsubsection{Rewriting the measure}
\label{holst_meas}

We have
\be
\rmd{\pi}^a_{IJ}=\rmd\pi^a_i\rmd\tilde{\pi}^a_i
\ee
The inverse of the relation between $\pi^a_i$ and $e^i_a$,
$\pi^a_i = \half \epsilon^{abc} \epsilon_{ijk} e^j_b e^k_c$, gives
\begin{equation}
\frac{\partial \pi^{a}_i}{\partial e^j_b} = \epsilon^{abc}
\epsilon_{ijk} e^k_c .
\end{equation}
Note $(ai)$ labels rows and $(bj)$ labels columns. Let $J^{ai}{}_{bj}$
denote this
matrix.
From the \textit{singular value decomposition theorem}, there exist
orthogonal matrices
$O^a{}_b$ and $O^i{}_j$ such that $O^b{}_a O^i{}_j e^j_b$ is diagonal,
that is
\begin{equation}
O^b{}_a O^i{}_j e^j_b = \lambda_a \delta^i_a .
\end{equation}
Let $O^{ai}{}_{bj} := O^a{}_b O^j{}_i$.  Then $O^{ai}{}_{bj}$ is also an
orthogonal matrix,
and we use it to define
\begin{equation}
\tilde{J}^{ai}{}_{bj} := O^{ai}{}_{bj} J^{bj}{}_{ck} O^{ck}{}_{dl}
= \sum_{c,k} \epsilon^{abc} \epsilon_{ijk} (\lambda_c \delta^k_c)
= \sum_c \epsilon^{abc} \epsilon_{ijc} \lambda_c
\end{equation}
where the symmetry of $\epsilon^{abc}$ and $\epsilon_{ijk}$ under
orthogonal transformations has been used.
From the above equation,
$\tilde{J}^{ai}{}_{bj}=0$ when $(i=j)$ or $(a=b)$ or $\{i,j\} \neq
\{a,b\}$.
From this one can deduce that,
for $i\neq a$, the row $(a,i)$ in $\tilde{J}^{ai}{}_{bj}$ has only one
non-zero element: the one in column
$(b=i,j=a)$. Reducing by minors along these 6 rows then gives
\begin{equation}
\label{detMt}
\det \tilde{J} = \tilde{J}^{12}{}_{21}\tilde{J}^{21}{}_{12}
\tilde{J}^{13}{}_{31} \tilde{J}^{31}{}_{13}
\tilde{J}^{23}{}_{32}\tilde{J}^{32}{}_{23} \det R
\end{equation}
where $R^i{}_j = \tilde{J}^{ii}{}_{jj}$ is the reduced $3 \times 3$
matrix.
The diagonal elements of $R$ are zero, so that $\det R$ has only 2
non-zero terms,
\begin{equation}
\label{detR}
\det R = \tilde{J}^{11}{}_{22} \tilde{J}^{22}{}_{33}
\tilde{J}^{33}{}_{11}
+ \tilde{J}^{11}{}_{33} \tilde{J}^{22}{}_{11} \tilde{J}^{33}{}_{22}
\end{equation}
As one can check, $\tilde{J}^{11}{}_{22} = \tilde{J}^{22}{}_{11} =
-\tilde{J}^{12}{}_{12} = -\tilde{J}^{21}{}_{21} = \lambda_3$, and
similarly for
cyclic permutations of $1,2,3$. Plugging this into (\ref{detR}) and then
(\ref{detMt}) gives
\begin{equation}
\det \tilde{J} = 2 (\lambda_1 \lambda_2 \lambda_3)^3 = 2 (\det e^i_a)^3.
\end{equation}
$\det J = \det \tilde{J}$, so that dropping the irrelevant $2$ factor,
\begin{equation}
\label{dBai}
\scrD \pi^{ai} = (\det e^k_c)^3 \scrD e^j_b .
\end{equation}

Next, define $\scrG^{ab} := f^i_a \tilde{\pi}^{bi}$, so that
\begin{equation}
\frac{\partial \scrG^{ab}}{\partial \tilde{\pi}^{ci}} = f^a_i \delta^b_c
.
\end{equation}
This is again block diagonal, whence
\begin{equation}
\det \left(\frac{\partial \scrG^{ab}}{\partial \tilde{\pi}^{ci}}\right)
= (\det f^a_i)^3 = (\det e^i_a)^{-3}
\end{equation}
so that
\begin{equation}
\label{d_tilBai}
\scrD \tilde{\pi}^{ai} = (\det e^i_a)^3 \scrD \scrG^{ab} =
(\det e^i_a)^3 \scrD \scrG^{(ab)} \scrD \scrG^{[ab]}
= (\det e^i_a)^3 \scrD S^{ab} \scrD e^0_b.
\end{equation}
Lastly, from (\ref{simpS}), $\scrD C^{ab} = (\det e^i_a)^6 \scrD
S^{ab}$, so that
\begin{equation}
\scrD \tilde{\pi}^{ai} = (\det e^i_a)^{-3} \scrD C^{ab} \scrD e^0_b.
\end{equation}

Coming to the lapse and shift, the Jacobian of the transformation
$(N,N^a) \mapsto e^I_t$ is
\be
\label{lapsh_jac}
J=\frac{\partial e_t^I}{\partial (N,N^a)}=\left(n^I,e_a^I\right)
\ee
On the other hand, the 4-volume element
\be
\label{lapsh_jac2}
\det
e_\a^I=\det\left(e_t^I,e_a^I\right)=\det\left(Nn_I+N^ae_{aI},e_a^I\right)=\det\left(Nn_I,e_a^I\right)=N\det
J
\ee
thus $|\det J|=V_s$ and $\rmd N\rmd N^a=\rmd e_t^I/V_s$.

Putting all the above relations together, we have
\begin{equation}
\label{holst_configmeas}
\scrD \pi^a_{IJ} \scrD N \scrD N^a = \frac{1}{V_s}
\scrD e^I_{\mu} \scrD C^{ab}.
\end{equation}
Note the above measure is $SO(\eta)$ covariant, consistent with the
$SO(\eta)$ covariance of the starting point.

\subsubsection{Final path integral}

Inserting (\ref{holst_configmeas}) into (\ref{holstcanint}), and
integrating out $C^{ab}$ finally gives
\be
\label{holst_final}
\scrZ&=&\int \scrD \conn_\mu^{IJ} \scrD e^I_\mu \ N^3 V_s^4\
\sqrt{|D_1|}\prod_{\a}\delta(\xi_\a)\ \exp i\int\ e^I\wedge
e^J\wedge\left(\dual
F_{IJ}-\frac{1}{\gamma}F_{IJ}\right)[\omega]\nonumber\\
&=&\int \scrD \conn_\mu^{IJ} \scrD e^I_\mu \ \cv^3 V_s\
\sqrt{|D_1|}\prod_{\a}\delta(\xi_\a)\ \exp i\int\ e^I\wedge
e^J\wedge\left(\dual
F_{IJ}-\frac{1}{\gamma}F_{IJ}\right)[\omega].\label{pi6}
\ee
Note that the integral in Eq.(\ref{pi6}) is restricted in the sector
$(II+)$. But if we want the integral to be over both the sectors $(II+)$
and $(II-)$, we will obtain
\be
\scrZ_{\pm}&=&\int_{II\pm} \scrD \conn_\mu^{IJ} \scrD e^I_\mu \ \cv^3
V_s\
\sqrt{|D_1|}\prod_{\a}\delta(\xi_\a)\ \cos\int\ e^I\wedge
e^J\wedge\left(\dual F_{IJ}-\frac{1}{\gamma}F_{IJ}\right)[\omega] .
\ee
In the existing spin-foam models in the literature
\cite{sfrevs, spinfoam1, spinfoam2}, sectors
(II+) and (II-) are not distinguished.  One can see from the above
equation, therefore, why it is generally the Cosine of the action and not the
exponential of the action that is expected to appear (and does
appear) in the asymptotic analysis of vertex amplitudes \cite{asym}, see also the discussions of the issue in some other different perspectives \cite{cos}

In the follows, we always use $\cz_\pm$ and $\int_{II\pm}$ to denote the
integral over both sectors, and $\cz$, $\int$ only to denote the
integral over a single sector $(II+)$.

\section{The construction of path-integral measure for Plebanski-Holst,
by way of Holst}
\label{pleb_sect}

In this section we would like to relate the previous Holst action
partition function with the partition function for the Plebanski-Holst
action. Our starting point for the reconstruction is Eq.(\ref{pi2}) (we
first only consider a single sector $II+$ for simplicity)
\begin{equation}
\scrZ=\int \scrD\conn_a^{IJ} \scrD\conn_t^{IJ} \scrD\pi_{IJ}^a \scrD N^a
\scrD N\
\delta(C_{\pi\pi}^{ab})\ N^3 V_s^5 \
\sqrt{|D_1|}\prod_{\a}\delta(\xi_\a)\ \exp i\int\rmd t \rmd^3x\
\left[\gdec{\pi}_{IJ}^a\dot{\conn}_a^{IJ} +
\conn_t^{IJ}G_{IJ}-N^aH_a-NH\right] \dummy
\end{equation}
where we use new notation for the simplicity constraint $C_{\pi\pi}^{ab}
:= C^{ab}$,
anticipating the introduction of further simplicity constraints.
To remind the reader,
\be
G_{IJ}&=& D_a \gdec{\pi}^a_{IJ} \nonumber\\
H_a&=&\frac{1}{2}F_{ab}^{IJ}[\conn]\ \gdec{\pi}^b_{IJ}\nonumber\\
H&=&\frac{1}{4\sqrt{\det q}}(F-\frac{1}{\g}*F)_{ab}^{IJ}[\conn]\
\pi^a_{IK}\ \pi^b_{JL}\ \eta^{KL}\nonumber\\
C_{\pi\pi}^{ab}&=& \eps^{IJKL}\pi^a_{IJ}\pi^b_{KL} .
\ee

\subsection{Basic strategy and some definitions}

In order to rewrite this path integral as a (generalized)
Plebanski path integral, one needs to change the variables
$\pi^a_{IJ}, N, N^a$ in favor of a constrained Plebanski two-form
$\plform^{IJ}_{\mu\nu}$.  If we define
$\prec Y, Z \succ := \frac{1}{4} \epsilon_{IJKL} Y^{IJ} Z^{KL}$,
the constraint on $\plform^{IJ}_{\mu\nu}$ is
\begin{equation}
\label{Bsimp}
\prec \plform_{\mu\nu}, \plform_{\rho\sigma} \succ = \frac{\scrV}{4!}
\epsilon_{\mu\nu\rho\sigma}
\end{equation}
where $\scrV := \epsilon^{\mu\nu\rho\sigma}\prec \plform_{\mu\nu},
\plform_{\rho\sigma} \succ$.
This constraint implies $X_{\mu\nu}^{IJ}$ takes one of the four forms
\begin{equation}
\label{plebsectors}
X_{\mu\nu}^{IJ} = \left\{\begin{array}{lc}
\pm 2 e^{[I}_\mu e^{J]}_{\nu} & (I\pm) \\
\pm \epsilon^{IJ}{}_{KL} e^K_\mu e^L_\nu & (II\pm)
 \end{array}\right.
\end{equation}
for some tetrad $e^I_\mu$. On-shell, $\scrV = \det e^I_\mu$, the
4-volume element.
Following \cite{bhnr}, we decompose $\plform^{IJ}_{\mu\nu}$ into
\begin{eqnarray}
\label{pidef}
\pi^a_{IJ} &:=& \half \epsilon^{abc} (\plform_{bc})_{IJ} \\
\label{betadef}
\beta_a^{IJ} &:=& \plform_{ta}^{IJ}
\end{eqnarray}
and (\ref{Bsimp}) becomes
\begin{eqnarray}
\nonumber
C_{\pi\pi}^{ab} &:=& \prec \pi^a, \pi^b \succ \approx 0 \\
\label{constr_comp}
(C_{\beta\beta})_{ab} &:=& \prec \beta_a, \beta_b \succ \approx 0 \\
\nonumber
(C_{\beta\pi})^b_a &:=& \text{traceless part of }
\prec \beta_a, \pi^b \succ \approx 0.
\end{eqnarray}
The first of these constraints was imposed in section \ref{holst_basic};
the four sectors
appearing there are the same four sectors here in (\ref{plebsectors}).
As in section \ref{config_holst},
we restrict to sector (II+).
The last two of the constraints (\ref{constr_comp}) are new.

As $\pi^a_{IJ}$ was coordinatized
by $e^I_a$ and the simplicity constraint $C_{\pi\pi}^{ab}$ in section
\ref{config_holst}, similarly
in this section we introduce coordinates for $\beta_a^{IJ}$.
Specifically, we will define a change of variables
\begin{equation}
\label{betachange}
\beta_a^{IJ} \leftrightarrow
(N, N^a, \tilde{C}_{\beta\beta}, \tilde{C}_{\beta\pi})
\end{equation}
where $\tilde{C}_{\beta\beta}$ and $\tilde{C}_{\beta\pi}$ have
the properties
\begin{enumerate}
\item If $C_{\pi\pi} = 0$, then $\tilde{C}_{\beta\pi} = C_{\beta\pi}$
\item If $\tilde{C}_{\beta\pi} = 0$, then
$\tilde{C}_{\beta\beta} = C_{\beta\beta}$.
\end{enumerate}
These properties will be used to replace $\tilde{C}_{\beta\pi},
\tilde{C}_{\beta\beta}$
in favor of $C_{\beta\pi}, C_{\beta\beta}$ in the final path integral in
section \ref{pleb_finalsect}.
$N$ and $N^a$ are defined as follows.  First define, in order,
\begin{enumerate}
\item $e^i_t := \half \epsilon^{ij}{}_{k} f^a_j \beta_a^{0k}$
\item $e^0_t := \frac{1}{3}f^a_i \left(\half \epsilon^i{}_{jk}
\beta^{jk}_a + e^0_a e^i_t\right)$
\end{enumerate}
where $f^a_i := (e_a^i)^{-1} = \left|\det \pi^b_{0j}\right|^{-\half}
\pi^a_{0i}$
and $e^0_a$ are as defined in \S\ref{config_holst}.
One can verify that when $\plform^{IJ}_{\mu\nu}$ is of the form (II+)
above,
this definition of $e^I_t$ coincides with the $t$ component of the
tetrad.
$N, N^a$ are then defined by the relation
\begin{equation}
\label{NNIdef}
e^I_t =: Nn^I + N^a e^I_a
\end{equation}
where recall from \S \ref{holst_basic} that $n^I$ is determined by
$\pi^a_{IJ}$ essentially via
$n^I \pi^a_{IJ} = 0$.

\subsection{The coordinate transformation and its bijectivity}

We begin by decomposing $\beta_a^{IJ}$ as
\begin{eqnarray}
\beta_a^i &:=& \beta_a^{0i} \\
\tilde{\beta}_a^i &:=& \half \epsilon^i{}_{jk} \beta_a^{jk} .
\end{eqnarray}
Then one sees that $e^i_t$ and $e^0_t$
\begin{eqnarray}
\label{eitbeta}
e^i_t &=& \half \epsilon^{ij}{}_k f^a_j \beta^k_a \\
e^0_t &=& \frac{1}{3} f^a_i (\tilde{\beta}^i_a + e^0_a e^i_t)
\end{eqnarray}
contain precisely the information
about the skew-symmetric part of $f^{ai} \beta^k_a$
and the trace of $f^a_i\tilde{\beta}^i_b$.
This leaves only the
symmetric part of $f^{ai} \beta^k_a$ and the trace-free part
of $f^a_i \tilde{\beta}^i_b$:
\begin{eqnarray}
C^{ij} &:=& f^{a(i} \beta^{j)}_a \\
K^a_b &:=& f^a_i \tilde{\beta}^i_b - \frac{1}{3} \delta_b^a f^c_j
\tilde{\beta}^j_c
\end{eqnarray}
so that there is a manifest isomorphism
\begin{equation}
\label{betaiso1}
\beta_a^{IJ} \equiv (\beta_a^i, \tilde{\beta}_a^i)
\leftrightarrow (e^i_t, e^0_t, C^{ij}, K^a_b).
\end{equation}
It is convenient to replace $K^a_b$ with a translation
depending only on $\pi^a_{IJ}, e^i_t$
\begin{equation}
\label{LKrel}
L^a_b = K^a_b + f^a_i e^0_b e^i_t - \frac{1}{3} \delta^a_b f^c_i e^0_c
e^i_t = \text{trace-free part of } f^a_i(\tilde{\beta}^i_b + e^0_b
e^i_t).
\end{equation}
\begin{equation}
\beta_a^{IJ}
\leftrightarrow (e^i_t, e^0_t, C^{ij}, L^a_b).
\end{equation}

We next invert the relations (\ref{eitbeta})-(\ref{LKrel}),
\begin{eqnarray}
\beta^i_a &=& \epsilon^i{}_{jk} e^k_t e^j_a + C^{ij}e_{aj} \\
\tilde{\beta}^i_a &=& e^i_b L^b_a +
2 e^{[i}_a e^{0]}_t
\end{eqnarray}
and substitute them into the expressions for
$C_{\beta\beta}, C_{\beta\pi}$.  This gives
\begin{eqnarray}
\label{Cbb}
\frac{s}{2}(C_{\beta\beta})_{ab} &=& \epsilon_{ijk} e^k_t e^i_c e^j_{(a}
L^c_{b)}
+ C_{ij} e^i_c e^j_{(a} L^c_{b)}
+ 2C_{ij} e^j_{(a} e^{[i}_{b)} e^{0]}_t
\\
\label{Cbpi}
s(C_{\beta\pi})^a_b &=&
\epsilon_{ijk} e^k_t e^j_b S^{ac} e^i_c
 + C_{ij} \epsilon^{acd} e_b^j e^i_c e^0_d
 + (\det e^k_d) L^a_b
 + C_{ij} e^j_b e^i_c S^{ac}
 - \frac{1}{3} \delta^a_b C_{ij} e^j_c e^i_d S^{cd}
\end{eqnarray}
where $S^{ab}$ is determined by $\pi^a_{IJ}$ as in section
\ref{holst_bij}.
Dropping the $S^{ab}$ terms from the second of these equations leads to
the definition of $\tilde{C}_{\beta\pi}$:
\begin{equation}
\label{Cbpi_t}
s(\tilde{C}_{\beta\pi})^a_b =
C_{ij} \epsilon^{acd} e_b^j e^i_c e^0_d
 + (\det e^k_d) L^a_b.
\end{equation}
So that $\tilde{C}_{\beta\pi} = C_{\beta\pi}$ when $C_{\pi\pi} = 0$.
With $e^i_t, e^0_t, C^{ij}$ fixed,
$L^a_b \mapsto (\tilde{C}_{\beta\pi})^a_b$ is manifestly bijective,
so that, composing with (\ref{betaiso1}),
\begin{equation}
\beta_a^{IJ} \leftrightarrow (e^i_t, e^0_t, C^{ij},
\tilde{C}_{\beta\pi})
\end{equation}
is bijective.

Solving (\ref{Cbpi}) for $L^a_b$ and substituting it into
(\ref{Cbb}) gives, after some manipulation,
\begin{equation}
\frac{s}{2} (C_{\beta\beta})_{ab}
= (\det e^k_d)^{-1}(\epsilon_{ijk} e^k_t + C_{ij})e^i_c e^j_{(a}
(\tilde{C}_{\beta\pi})^c_{b)}
+ (e^0_t - e^k_t f_k^e e^0_e) e^i_a e^j_b C_{ij}.
\end{equation}
Dropping the $\tilde{C}_{\beta\pi}$ term leads to the definition
\begin{equation}
\label{Cbb_t}
\frac{s}{2} \tilde{C}_{\beta\beta} :=
(e^0_t - e^k_t f_k^e e^0_e) e^i_a e^j_b C_{ij}
\end{equation}
so that $\tilde{C}_{\beta\beta} = C_{\beta\beta}$ when
$\tilde{C}_{\beta\pi} = 0$.
In order to understand the significance of the prefactor in
(\ref{Cbb_t}), we prove the following
\begin{lem}
\label{vol4lem}
$(\det e^i_a)(e^0_t - e^0_h f^h_k e^k_t)$ is equal to the 4-volume
element
$\scrV = \det e^I_\alpha$.
\end{lem}
{\startproof
\begin{eqnarray*}
\det e^I_\alpha &=& \frac{1}{4!} \epsilon^{\alpha \beta \gamma \delta}
\epsilon_{IJKL} e^I_\alpha e^J_\beta e^K_\gamma e^L_\delta \\
&=& \frac{1}{3!} \epsilon^{tabc} \epsilon_{IJKL} e^I_t e^J_a e^K_b e^L_c
= \frac{1}{3!} \epsilon^{abc} \epsilon_{IJKL} e^I_t e^J_a e^K_b e^L_c \\
&=& \frac{1}{3!} \epsilon^{abc} \epsilon_{jkl} e^0_t e^j_a e^k_b e^l_c
+ \half \epsilon^{abc} \epsilon_{i0kl} e^i_t e^0_a e^k_b e^l_c \\
&=& (\det e^i_a) e^0_t - e^i_t e^0_a (\det e^j_b) f^a_i \\
&=& (\det e^i_a)(e^0_t - e^i_t f^a_i e^0_a) .
\end{eqnarray*}
\finishproof}
By assumption $\scrV$ and $\det e^i_a$ are non-zero, so the prefactor in
(\ref{Cbb_t}) is non-zero, whence, from (\ref{Cbb_t}),
$C^{ij} \mapsto (\tilde{C}_{\beta\beta})_{ab}$
is manifestly bijective.  Thus
\begin{equation}
\beta_a^{IJ} \leftrightarrow (e^i_t, e^0_t, \tilde{C}_{\beta\beta},
\tilde{C}_{\beta\pi})
\end{equation}
is a bijection.

Finally,
\begin{equation}
e^I_t \leftrightarrow (N, N^a)
\end{equation}
as defined in (\ref{NNIdef}) is clearly an isomorphism.  Putting these
together, we see that
\begin{displaymath}
\beta_a^{IJ} \leftrightarrow (N, N^a, \tilde{C}_{\beta\beta},
\tilde{C}_{\beta\pi})
\end{displaymath}
is a bijection as claimed.

\subsection{The change of measure}

In this subsection, we calculate the measure $\scrD \beta_a^{IJ}$ in
terms of
$\scrD N \scrD N^a \scrD \tilde{C}_{\beta\beta} \scrD
\tilde{C}_{\beta\pi}$.
First,
\begin{equation}
\label{bbbt}
\scrD \beta_a^{IJ} = \scrD \beta_a^i \scrD \tilde{\beta}_a^i .
\end{equation}
We next wish to relate $\scrD \beta^k_a$ and $\scrD e^i_t \scrD C_{jk}$.
Our strategy is to first define
\begin{equation}
\scrF_{ij} := \beta_{ai} f^a_j ,
\end{equation}
find the relation between $\scrD \beta_{ai}$ and $\scrD \scrF_{ij}$,
and then use $\scrD \scrF_{ij} = \scrD \scrF_{(ij)} \scrD \scrF_{[ij]}$.
As $C_{ij} = \scrF_{(ij)}$ and $e^j_t = \half \epsilon^{jkl}
\scrF_{[kl]}$,
in fact this becomes $\scrD \scrF_{ij} = \scrD C_{ij} \scrD e^j_t$.
First,
\begin{equation}
\frac{\partial \scrF_{ij}}{\partial \beta_{ak}} = \delta^k_i f^a_j .
\end{equation}
This is a block diagonal matrix, with three blocks each equal to
$f^a_i$:
\begin{equation}
\det \frac{\partial \scrF_{ij}}{\partial \beta_{ak}}
= (\det f^a_j)^3 = (\det e^j_a)^{-3}
\end{equation}
so that
\begin{equation}
\label{dbai}
\scrD \beta_{ai} = (\det e^j_a)^3 \scrD \scrF_{ij} = (\det e^j_a)^3
\scrD C_{ij} \scrD e^k_t .
\end{equation}

Next, to rewrite $\scrD \tilde{\beta}^i_a$, first define
$\scrH^a_b := f^a_i \tilde{\beta}^i_b$. Then
\begin{equation}
\frac{\partial \scrH^a_b}{\partial \tilde{\beta}^i_c} = f^b_j \delta^c_a
\end{equation}
so that the matrix is block diagonal with three blocks, each equal to
$f^a_j$:
\begin{equation}
\det \left( \frac{\partial \scrH^b_a}{\partial \tilde{\beta}^i_c}
\right)
= (\det f^a_j)^3 = (\det e^j_a)^{-3},
\end{equation}
so that
\begin{equation}
\scrD \tilde{\beta}^k_a = (\det e^j_c)^3 \scrD \scrH^a_b
= (\det e^j_a)^3 \scrD K^a_b \scrD \tr \scrH
\end{equation}
where we have used that $K^a_b$ is the traceless part
of $\scrH^a_b$.  As $L^a_b$ is just a translation of $K^a_b$
(\ref{LKrel}) by
a term involving only
$e^i_t$, which is being held constant, one can replace
$\scrD K^a_b$ with $\scrD L^a_b$.  Likewise $e^0_t$ is
just a translation of $\tr \scrH$ by a term involving only
$e^i_t$, so that $\scrD \tr \scrH = \scrD e^0_t$, whence
\begin{equation}
\label{dtbia}
\scrD \tilde{\beta}^i_a = (\det e^j_a)^3 \scrD L^a_b \scrD e^0_t .
\end{equation}
Putting together (\ref{bbbt}, \ref{dbai}, \ref{dtbia}),
\begin{equation}
\label{dBIJ_2}
\scrD \beta_a^{IJ} = (\det e^j_a)^{6} \scrD e^I_t \scrD C_{ij} \scrD
L^a_b.
\end{equation}

To perform the change of variables
$(C_{ij}, L^a_b) \rightarrow (\tilde{C}_{\beta\beta},
\tilde{C}_{\beta\pi})$,
we proceed in two steps, $(C_{ij},L^a_b) \rightarrow (C_{ij},
\tilde{C}_{\beta\pi})
\rightarrow (\tilde{C}_{\beta\beta}, \tilde{C}_{\beta\pi})$.
Holding $C_{ij}$ constant, the Jacobian of the first transformation
is
\begin{equation}
\frac{\partial (\tilde{C}_{\beta\pi})^b_a}{\partial L^c_d} \approx (\det
e^k_d)
(\delta^c_a \delta^b_d - \frac{1}{3}\delta^b_a \delta^c_d)
\end{equation}
Here $\delta^c_a \delta^b_d - \frac{1}{3}\delta^b_a \delta^c_d$ is the
identity matrix
on the space of trace-less matrices.  The space of trace-less matrices
is 8-dimensional,
so that
\begin{equation}
\det\left(\frac{\partial (\tilde{C}_{\beta\pi})^b_a}{\partial
L^i_j}\right) \approx (\det e^k_d)^8
\end{equation}
whence
\begin{equation}
\label{dLij}
\scrD L^a_b = (\det e^k_d)^{-8} \scrD \tilde{C}_{\beta\pi} .
\end{equation}
Next, to replace $C_{ij}$ in favor of $C_{\beta\beta}$. We want the
Jacobian of the matrix
\begin{equation}
\frac{\partial (\tilde{C}_{\beta\beta})_{ab}}{\partial C_{ij}}
= 2s (e^0_t - e^0_h f^h_k e^k_t) e^{(i}_a e^{j)}_b ,
\end{equation}
that is,
\begin{equation}
\det \left( \frac{\partial (\tilde{C}_{\beta\beta})_{ab}}{\partial
C_{ij}} \right)
= 2^6(e^0_t - e^0_h f^h_k e^k_t)^6 \det_{(ab,ij)} e^{(i}_a e^{j)}_b .
\end{equation}
Let $H^{ab}{}_{ij} := e^{(i}_a e^{j)}_b$.  Recall from
\S\ref{holst_meas}
the orthogonal matrices $O^b{}_a$ and $O^i{}_j$
satisfying $O^b{}_a O^i{}_j e^j_b = \lambda_a \delta^i_a$; let
$S^{ab}{}_{cd} := O^{(a}{}_c O^{b)}{}_d$ and $S^{ij}{}_{kl}:= O^k{}_{(i}
O^l{}_{j)}$,
orthogonal matrices. Then
\begin{equation}
\tilde{H}^{ab}{}_{ij} := S^{ab}{}_{cd} H^{cd}{}_{kl} S^{kl}{}_{ij}
= \lambda_a \lambda_b \delta^{(i}_a \delta^{j)}_b
\end{equation}
which is a diagonal 6 by 6 matrix.
\begin{equation}
\det \tilde{H} =
(\lambda_1^2)(\lambda_2^2)(\lambda_3^2)(\lambda_1\lambda_2)(\lambda_1
\lambda_3)(\lambda_2 \lambda_3)
= (\det e^i_a)^4
\end{equation}
so
\begin{equation}
\label{dCij}
\scrD C_{ij} = (e^0_t - e^0_h f^h_k e^k_t)^{-6} (\det e^i_a)^{-4} \scrD
\tilde{C}_{\beta\beta}
\end{equation}
where an irrelevant overall numerical coefficient was dropped.

Finally one performs the change of variables $e_t^I \rightarrow
(N,N^a)$.
Using (\ref{lapsh_jac}, \ref{lapsh_jac2}) from the last section, and
(\ref{dLij},\ref{dCij},\ref{dBIJ_2}), one has finally
\begin{equation}
\scrD \beta_a^{IJ} = (e^0_t - e^0_c f^c_k e^k_t)^{-6} (\det e^j_a)^{-6}
V_s
\scrD N \scrD N^a \scrD \tilde{C}_{\beta\beta} \scrD
\tilde{C}_{\beta\pi} .
\end{equation}
With lemma \ref{vol4lem} this becomes
\begin{equation}
\label{final_dBIJ}
\scrD \beta_a^{IJ} = \scrV^{-6} V_s
\scrD N \scrD N^a \scrD \tilde{C}_{\beta\beta} \scrD
\tilde{C}_{\beta\pi}  .
\end{equation}

\subsection{Final path integral}
\label{pleb_finalsect}

Starting with the canonical path integral (\ref{holstcanint}), and
inserting
$\int \scrD \tilde{C}_{\beta\pi} \scrD \tilde{C}_{\beta\beta}
\delta(\tilde{C}_{\beta\pi})
\delta(\tilde{C}_{\beta\beta}) = 1$, we have
\be
\scrZ&=&\int \scrD \conn_\mu^{IJ} \scrD \pi_{IJ}^a \scrD N^a \scrD N
\scrD \tilde{C}_{\beta\pi}
\scrD \tilde{C}_{\beta\beta}
\delta(C_{\pi\pi})
\delta(\tilde{C}_{\beta\pi})\delta(\tilde{C}_{\beta\beta})N^3 V_s^5 \exp
i\int\rmd t \rmd^3x\
\left[\gdec{\pi}_{IJ}^a\dot{\conn}_a^{IJ}-\conn_t^{IJ}G_{IJ}-N^aH_a-NH\right]\nonumber\\
&&\times\ \sqrt{|D_1|}\prod_{\a}\delta(\xi_\a).
\ee
Using (\ref{final_dBIJ}) then gives
\begin{equation}
\scrZ = \int \scrD \conn_\mu^{IJ} \scrD \pi^a_{IJ} \scrD \beta_a^{IJ}
\delta(C_{\pi\pi})
\delta(\tilde{C}_{\beta\pi})\delta(\tilde{C}_{\beta\beta}) N^9 V_s^{10}
\ \sqrt{|D_1|}\prod_{\a}\delta(\xi_\a)\
\exp i\int\rmd t \rmd^3x\
\left[\gdec{\pi}_{IJ}^a\dot{\conn}_a^{IJ}-\conn_t^{IJ}G_{IJ}-N^aH_a
-NH\right].
\end{equation}
We next use the presence of $\delta(\tilde{C}_{\beta\pi})$ and the fact
that this
enforces
$\tilde{C}_{\beta\beta} = C_{\beta\beta}$ to
replace $\tilde{C}_{\beta\beta}$ in favor of $C_{\beta\beta}$; then we
use
the presence of $\delta(C_{\pi\pi})$ and the fact that it enforces
$\tilde{C}_{\beta\pi} = C_{\beta\pi}$ to replace
$\tilde{C}_{\beta\pi}$ in favor of $C_{\beta\pi}$, yielding
\begin{equation}
\label{plebpath_mid}
\scrZ = \int \scrD \conn_\mu^{IJ} \scrD \pi^a_{IJ} \scrD \beta_a^{IJ}
\delta(C_{\pi\pi}) \delta(C_{\beta\pi})\delta(C_{\beta\beta}) N^9
V_s^{10}
\exp i\int\rmd t \rmd^3x\
\left[\gdec{\pi}_{IJ}^a\dot{\conn}_a^{IJ}-\conn_t^{IJ}G_{IJ}-N^aH_a
-NH\right].
\end{equation}
\begin{lem}
When the simplicity constraints are satisfied,
\begin{equation}
\label{toprove}
\plform_{ta}^{IJ} =  \half N^c \epsilon_{abc} \pi^{bIJ}
- \frac{N}{\sqrt{\det q}} \epsilon_{abc}\pi^{bI}{}_K \pi^{cKJ}
\end{equation}
\end{lem}
{\startproof
With the simplicity constraints satisfied,
\begin{displaymath}
\plform_{\mu\nu}^{IJ} = \epsilon^{IJ}{}_{KL} e^K_\mu e^L_\nu
\end{displaymath}
so that
\begin{eqnarray}
\nonumber
\plform_{ta}^{IJ} &=& \epsilon^{IJ}{}_{KL} e^K_t e^L_a \\
\nonumber
&=&  \epsilon^{IJ}{}_{KL} (N^b e_b^K + Nn^K) e^L_a \\
\nonumber
&=& N^b \plform_{bc}^{IJ} + N n^K \epsilon_K{}^{IJ}{}_L e^L_a\\
\label{lem_lhs}
&=& \half N^b \epsilon_{acb} \pi^{bIJ} + N n^K \epsilon_K{}^{IJ}{}_L
e^L_a.
\end{eqnarray}
The first term here matches the first term on the right hand side of
(\ref{toprove}).
The second term on the right hand side of (\ref{toprove}) is
\begin{eqnarray}
\nonumber
\frac{-N}{\sqrt{\det q}} \epsilon_{abc}\pi^{bI}{}_K \pi^{cKJ}
&=& \frac{-N}{4\sqrt{\det q}} \epsilon_{abc} \epsilon^{bde}
\epsilon^{cfg}
\epsilon^I{}_{KMN} \epsilon^{KJ}{}_{PQ}e_d^M e_e^N e_f^P e_g^Q \\
\label{secterm}
&=& \frac{-N}{2 \sqrt{\det q}} \epsilon^{bde}
\epsilon^I{}_{KMN} \epsilon^{KJ}{}_{PQ}e_d^M e_e^N e_a^P e_b^Q.
\end{eqnarray}
Now,
\begin{eqnarray*}
\epsilon^{bde} e_b^Q e_d^M e_e^N
= \epsilon^{tbde} e_b^Q e_d^M e_e^N
= \scrV \eta^{tbde} e_b^Q e_d^M e_e^N
= \scrV e^t_R \epsilon^{RQMN}
\end{eqnarray*}
where $\eta^{\mu\nu\rho\sigma}$ is the inverse volume form, and
$e^t_R$ is the indicated component of $e^\mu_I := (e^I_\mu)^{-1}$.
But $e^t_R = (\partial_\mu t) e^\mu_R = \frac{-1}{N} n_\mu e^\mu_R =
\frac{-1}{N} n_R$,
so that
\begin{equation}
\epsilon^{bde} e_b^Q e_d^M e_e^N
= -\sqrt{\det q} n_R \epsilon^{RQMN}.
\end{equation}
Thus (\ref{secterm}) becomes
\begin{eqnarray*}
\frac{-N}{\sqrt{\det q}} \epsilon_{abc}\pi^{bI}{}_K \pi^{cKJ}
&=&  \frac{N}{2} \epsilon^{bde}
\epsilon^I{}_{KMN} \epsilon^{KJ}{}_{PQ} n_R \epsilon^{RQMN} e_a^P \\
&=& N n^K \epsilon_K{}^{IJ}{}_P e^P_a
\end{eqnarray*}
matching the second term in (\ref{lem_lhs}) and completing the
proof.\finishproof}

\begin{cor}
On-shell with respect to the simplicity constraints,
\begin{displaymath}
\int \rmd t \rmd^3 x
\left[\gdec{\pi}_{IJ}^a\dot{\conn}_a^{IJ}+\conn_t^{IJ}G_{IJ}-N^aH_a-NH\right]
=
\int (\plform - \frac{1}{\gamma} \dual \plform)_{IJ} \wedge F^{IJ}
=:
\int \plconj_{IJ} \wedge F^{IJ}
\end{displaymath}
\end{cor}
{\startproof
\begin{eqnarray*}
\int \plconj_{IJ} \wedge F^{IJ}
&=&
\frac{1}{4} \int \rmd t \rmd^3 x \epsilon^{\mu\nu\rho\sigma}
\plconj_{\mu\nu IJ} F^{IJ}_{\rho\sigma}
= \half \int \rmd t \rmd^3 x \left[\epsilon^{abc} \plconj_{abIJ}
\partial_t \conn_c^{IJ}
+ \conn_t^{IJ} (D_c \plconj_{ab})_{IJ} \epsilon^{abc}
+ \epsilon^{abc} \plconj_{tcIJ} F_{ab}^{IJ}\right] \\
&=& \int \rmd t \rmd^3 x \left[ \gdec{\pi}^c_{IJ} \dot{\conn}_c^{IJ}
+ \conn_t^{IJ} \gdec{\pi}^c_{IJ}
+ \half \epsilon^{abc} (\plform - \frac{1}{\gamma} \plform)_{tcIJ}
F_{ab}^{IJ} \right] \\
&=& \int \rmd t \rmd^3 x \left[ \gdec{\pi}^c_{IJ} \dot{\conn}_c^{IJ}
+ \conn_t^{IJ} \gdec{\pi}^c_{IJ}
+ \frac{1}{4} \epsilon^{abc} N^e \epsilon_{cde} \gdec{\pi}^d_{IJ}
F_{ab}^{IJ}
- \frac{N}{4\sqrt{\det q}} \epsilon^{abc} \epsilon_{cde}
\left[ \pi^d \pi^e  - \frac{1}{\gamma} \dual (\pi^d \pi^e)\right]_{IJ}
F^{IJ}_{ab} \right] \\
&=& \int \rmd t \rmd^3 x \left[ \gdec{\pi}^c_{IJ} \dot{\conn}_c^{IJ}
+ \conn_t^{IJ} G_{IJ}
- N^a H_a - N H \right]
\end{eqnarray*}
\finishproof}
Substituting this into the path integral (\ref{plebpath_mid}), one has
finally
\be
\label{plebpath_final}
\cz&=&\int \scrD \conn_\mu^{IJ} \scrD X^{IJ}_{\mu\nu} \delta(C_{\pi\pi})
\delta(C_{\beta\pi}) \delta(C_{\beta\beta}) N^9 V_s^{10}\
\sqrt{|D_1|}\prod_{\a}\delta(\xi_\a)\ \exp i  \int\ (\plform -
\frac{1}{\gamma} \dual \plform)_{IJ} \wedge F^{IJ}\nonumber\\
&=&\int \scrD \conn_\mu^{IJ} \scrD X^{IJ}_{\mu\nu} \delta(C_{\pi\pi})
\delta(C_{\beta\pi}) \delta(C_{\beta\beta}) \cv^9 V_s\
\sqrt{|D_1|}\prod_{\a}\delta(\xi_\a)\ \exp i  \int\ (\plform -
\frac{1}{\gamma} \dual \plform)_{IJ} \wedge F^{IJ}.
\ee
Note that this integral is restricted to the solution sector $(II+)$ of
the simplicity constraint. We can extend the integral to include both
sectors $(II+)$ and $(II-)$ without changing the form of the integrand,
i.e. we obtain
\be
\cz_\pm&=&\int_{II\pm} \scrD \conn_\mu^{IJ} \scrD X^{IJ}_{\mu\nu}
\delta(C_{\pi\pi}) \delta(C_{\beta\pi}) \delta(C_{\beta\beta}) \cv^9
V_s\
\sqrt{|D_1|}\prod_{\a}\delta(\xi_\a)\ \exp i  \int\ (\plform -
\frac{1}{\gamma} \dual \plform)_{IJ} \wedge F^{IJ}.
\ee
In the following subsection, we show another way to construct the
Plebanski-Holst path-integral from the Holst path-integral, where we
implement both sectors $(II+)$ and $(II-)$.

\subsection{An alternative way to construct Plebanski-Holst
path-integral from Holst}

In this subsection we would like to give another derivation from the the
Holst action partition function to the partition of Plebanski-Holst
action. Such a derivation is made by transforming delta functions in the
integral. Our starting point for this alternative derivation is also
Eq.(\ref{pi2}), but with a integral over both sectors $(II+)$ and
$(II-)$
\be
\cz_\pm=\int_{II\pm}\cd\o_a^{IJ}\cd\o_t^{IJ}\cd\pi_{IJ}^a\cd N^a\cd N\
\delta(C^{ab})\ \cv^3 V_s^2 \sqrt{|D_1|}\prod_{\a}\delta(\xi_\a)\ \exp
i\int\rmd t\int\rmd^3x\
\left[\gdec{\pi}_{IJ}^a\dot{\o}_a^{IJ}-\o_t^{IJ}G_{IJ}-N^aH_a+NH\right]
\ee
We define a new tensor field $X_{tc}^{IJ}$ by
\be
X_{tc}^{IJ}:=\frac{1}{2}\eps_{abc}\ N^a\ \pi^{bIJ}-\frac{N}{\sqrt{\det
q}}\ \eps_{abc}\ \pi^{aIK}\ \pi^{bJL}\ \eta_{KL}
\ee
then the action on the exponential is again expressed as a BF action as
it is shown above
\be
S:=\int\rmd t\int\rmd^3x\
\left[\gdec{\pi}_{IJ}^a\dot{\o}_a^{IJ}-\o_t^{IJ}G_{IJ}-N^aH_a+NH\right]
=
\int (\plform - \frac{1}{\gamma} \dual \plform)_{IJ} \wedge F^{IJ}
=:
\int \plconj_{IJ} \wedge F^{IJ}
\ee
Therefore in terms of the new field $X_{tc}^{IJ}$ we can re-express the
path-integral as a constrained BF theory:
\be
\cz_\pm&=&\int_{II\pm}\cd\o_a^{IJ}\cd\o_t^{IJ}\cd\pi_{IJ}^a\cd
X_{tc}^{IJ}\cd N^a\cd N\ \cv^3
V_s^2\sqrt{|D_1|}\prod_{\a}\delta(\xi_\a)\ \exp
i \int (\plform - \frac{1}{\gamma} \dual \plform)_{IJ} \wedge F^{IJ}
\nonumber\\
&&\times\delta^6\left(\eps^{IJKL}\pi^a_{IJ}\pi^b_{KL}\right)\
\delta^{18}\left(X_{tc}^{IJ}+\frac{1}{2}\eps_{abc}\ N^a\
\pi^{bIJ}-\frac{N}{4\sqrt{\det q}}\ \eps_{abc}\ \pi^{aIK}\ \pi^{bJL}\
\eta_{KL}\right)
\ee
As a first step for recovering the full Plebanski simplicity
constraints, we divide the 18 $\delta$-functions into two collections,
each of which has 9 $\delta$-functions. Then we transform the two
collections of $\delta$-functions in two different way, i.e.
\be
&&\delta^{18}\left(X_{tc}^{IJ}+\frac{1}{2}\eps_{abc}\ N^a\
\pi^{bIJ}-\frac{N}{4\sqrt{\det q}}\ \eps_{abc}\ \pi^{aIK}\ \pi^{bJL}\
\eta_{KL}\right)\nonumber\\
&=&\delta^{9}\left(X_{tc}^{0i}+\frac{1}{2}\eps_{abc}\ N^a\
\pi^{b0i}-\frac{N}{4\sqrt{\det q}}\ \eps_{abc}\ \pi^{a0l}\
\pi^{bil}\right)\ \delta^9\left(X_{tc}^{ij}+\frac{1}{2}\eps_{abc}\ N^a\
\pi^{bij}-\frac{N}{4\sqrt{\det q}}\ \eps_{abc}\ \pi^{aiK}\ \pi^{bjL}\
\eta_{KL}\right)\nonumber\\
&=&\left(\text{Jacobian}\right)\nonumber\\
&&\times\delta^{9}\left(X_{tc}^{0i} X_{td}^{jk} \eps_{ijk}+X_{tc}^{ij}
X_{td}^{0k} \eps_{ijk}+\frac{1}{2}N^a\eps_{abc}\left[\pi^{b0i}
X_{td}^{jk}+\pi^{bij}X_{td}^{0k} \right]\eps_{ijk}-\frac{N}{4\sqrt{\det
q}}\eps_{abc}\left[\pi^{a0l} \pi^{bil} X_{td}^{jk}+ \pi^{aiK} \pi^{bjL}
\eta_{KL}X_{td}^{0k}\right]\eps_{ijk}\right)\nonumber\\
&&\times \delta^9\left(X_{tc}^{ij}\
\pi^{d0k}\eps_{ijk}+X_{tc}^{0i}\pi^{djk}\eps_{ijk}+\frac{1}{2}\eps_{abc}
N^a\left[\pi^{bij}\pi^{d0k}+\pi^{b0i}\pi^{djk}\right]\eps_{ijk}-\frac{N}{4\sqrt{\det
q}}\eps_{abc}\left[\pi^{aiK}\ \pi^{bj}_{\ \ K}\pi^{d0k}+\pi^{a0l}\
\pi^{bil}\pi^{djk}\right]\eps_{ijk}\right)\nonumber
\ee
We define some notations:
\be
\chi_c^i&:=&X_{tc}^{0i}+\frac{1}{2}\eps_{abc}\ N^a\
\pi^{b0i}-\frac{N}{4\sqrt{\det q}}\ \eps_{abc}\ \pi^{a0l}\
\pi^{bil}\nonumber\\
\tilde{\chi}_c^{ij}&:=&X_{tc}^{ij}+\frac{1}{2}\eps_{abc}\ N^a\
\pi^{bij}-\frac{N}{4\sqrt{\det q}}\ \eps_{abc}\ \pi^{aiK}\ \pi^{bjL}\
\eta_{KL}\nonumber\\
\cf_{cd}&:=&X_{tc}^{0i} X_{td}^{jk} \eps_{ijk}+X_{tc}^{ij} X_{td}^{0k}
\eps_{ijk}+\frac{1}{2}N^a\eps_{abc}\left[\pi^{b0i}
X_{td}^{jk}+\pi^{bij}X_{td}^{0k} \right]\eps_{ijk}-\frac{N}{4\sqrt{\det
q}}\eps_{abc}\left[\pi^{a0l} \pi^{bil} X_{td}^{jk}+ \pi^{aiK} \pi^{bjL}
\eta_{KL}X_{td}^{0k}\right]\eps_{ijk}\nonumber\\
&=&\chi_c^iX_{td}^{jk}\eps_{ijk}+\tilde{\chi}_c^{ij}X_{td}^{0k}\eps_{ijk}\nonumber\\
\tilde{\cf}_{c}^d&:=&X_{tc}^{ij}\
\pi^{d0k}\eps_{ijk}+X_{tc}^{0i}\pi^{djk}\eps_{ijk}+\frac{1}{2}\eps_{abc}
N^a\left[\pi^{bij}\pi^{d0k}+\pi^{b0i}\pi^{djk}\right]\eps_{ijk}-\frac{N}{4\sqrt{\det
q}}\eps_{abc}\left[\pi^{aiK}\ \pi^{bj}_{\ \ K}\pi^{d0k}+\pi^{a0l}\
\pi^{bil}\pi^{djk}\right]\eps_{ijk}\nonumber\\
&=&\tilde{\chi}_c^{ij}\pi^{d0k}\eps_{ijk}+\chi_c^i\pi^{djk}\eps_{ijk}
\ee
then the Jacobian from above $\delta$-function transformation is the
determinant of the transformation matrix on the constraint surface
\be
\frac{\partial (\cf_{cd}, \tilde{\cf}^d_c)}{\partial (\chi^{i}_{a},
\tilde{\chi}^{jk}_{a}) }=
\begin{pmatrix}
      &\delta^a_c\
\left(X_{td}^{jk}+\tilde{\chi}^{jk}_d\right)\eps_{ijk}\ \ \ ,&\
\delta^a_d\ \left(X_{tc}^{0i}+\chi^i_c\right)\eps_{ijk}\ \  \\
      &\delta^a_c\ \pi^{djk}\eps_{ijk}\ \ \ ,  &\  \delta^a_c\
\pi^{d0i}\eps_{ijk}\ \
\end{pmatrix}\approx
\begin{pmatrix}
      &\delta^a_c\ X_{td}^{jk}\eps_{ijk}\ \ \ ,&\ \delta^a_d\
X_{tc}^{0i}\eps_{ijk}\ \  \\
      &\delta^a_c\ \pi^{djk}\eps_{ijk}\ \ \ ,  &\  \delta^a_c\
\pi^{d0i}\eps_{ijk}\ \
\end{pmatrix}.
\ee
We can see that
\be
\det\Big[\begin{pmatrix}
      &\delta^a_c\ X_{td}^{jk}\eps_{ijk}\ \ \ ,&\ \delta^a_d\
X_{tc}^{0i}\eps_{ijk}\ \  \\
      &\delta^a_c\ \pi^{djk}\eps_{ijk}\ \ \ ,  &\  \delta^a_c\
\pi^{d0i}\eps_{ijk}\ \
\end{pmatrix}
\Big]=\left(\det\Big[\begin{pmatrix}
      &\ X_{td}^{jk}\eps_{ijk}\ \ \ ,&\ \ X_{tc}^{0i}\eps_{ijk}\ \  \\
      &\ \pi^{djk}\eps_{ijk}\ \ \ ,  &\   \pi^{d0i}\eps_{ijk}\ \
\end{pmatrix}\Big]\right)^3=\left(\det[X_{\a\b}^{IJ}]\right)^3=\cv^9
\ee
Therefore we insert back this result, and further divide the first
collection into its symmetric and anti-symmetric parts. After some
manipulation, we have
\be
&&\delta^{18}\left(X_{tc}^{IJ}+\frac{1}{2}\eps_{abc}\ N^a\
\pi^{bIJ}-\frac{N}{4\sqrt{\det q}}\ \eps_{abc}\ \pi^{aIK}\ \pi^{bJL}\
\eta_{KL}\right)\nonumber\\
&=&\cv^9\ \delta^{3}\left(\frac{1}{2}N^a\left[\eps_{ab[c} \pi^{b0i}
X_{td]}^{jk}+\eps_{ab[c} \pi^{bij}X_{td]}^{0k}
\right]\eps_{ijk}-\frac{N}{4\sqrt{\det q}}\left[\eps_{ab[c} \pi^{a0l}
\pi^{bil} X_{td]}^{jk}+\eps_{ab[c} \pi^{aiK} \pi^{bjL}
\eta_{KL}X_{td]}^{0k}\right]\eps_{ijk}\right)\nonumber\\
&&\times \delta^{6}\left(X_{tc}^{IJ} X_{td}^{KL}
\eps_{IJKL}+\frac{1}{2}N^a\left[\eps_{ab(c} \pi^{b0i}
X_{td)}^{jk}+\eps_{ab(d} \pi^{bjk}X_{tc)}^{0i}
\right]\eps_{ijk}-\frac{N}{4\sqrt{\det q}}\left[\eps_{ab(c} \pi^{a0l}
\pi^{bil} X_{td)}^{jk}+\eps_{ab(d} \pi^{ajK} \pi^{bkL}
\eta_{KL}X_{tc)}^{0i}\right]\eps_{ijk}\right)\nonumber\\
&&\times \delta^9\left(\frac{1}{2}X_{tc}^{IJ}\
\pi^{dKL}\eps_{IJKL}-\frac{N}{4\sqrt{\det q}}\eps_{abc}\left[\pi^{aiK}\
\pi^{bj}_{\ \ K}\pi^{d0k}+\pi^{a0l}\
\pi^{bil}\pi^{djk}\right]\eps_{ijk}\right)\label{deltas}
\ee
Since the simplicity constraint
$C^{ab}=\eps^{IJKL}\pi^a_{IJ}\pi^b_{KL}=0$ implies that there exists a
non-degenerated $so(3)$-valued one form $e_a^i$ and an another
independent 1-form $e^0_a$ such that (we are working in both two sectors
$(II\pm)$) $\pi^a_{IJ}=\pm\eps^{abc}e_b^Ke_c^L\eps_{IJKL}$. Then we
obtain the following lemma:

\begin{lem}
On the constraint surface defined by the delta function
$\delta(C^{ab})$, the field $X_{tc}^{IJ}$ can be written as
\be
X_{tc}^{IJ}=\pm e_t^Ke_c^L\eps^{IJ}_{\ \ KL}=\pm
(Nn^K+N^ae_a^K)e_c^L\eps^{IJ}_{\ \ KL}
\ee
where $e_\a^I$ $\a=t,1,2,3$ form a non-degenerate tetrad field for
non-vanished $N$. Thus the 18 delta functions
\be
\delta^{18}\left(X_{tc}^{IJ}+\frac{1}{2}\eps_{abc}\ N^a\
\pi^{bIJ}-\frac{N}{4\sqrt{|\det q|}}\ \eps_{abc}\ \pi^{aIK}\ \pi^{bJL}\
\eta_{KL}\right)
\ee
can essentially be written as
$\delta^{18}\left(X_{tc}^{IJ}-e_t^Ke_c^L\eps^{IJ}_{\ \
KL}\right)+\delta^{18}\left(X_{tc}^{IJ}+e_t^Ke_c^L\eps^{IJ}_{\ \
KL}\right)$.
\end{lem}

{\startproof The lemma follows staight-forwardly from the definition of
$X^{IJ}_{tc}$:
\be
X_{tc}^{IJ}=\frac{1}{2}\eps_{abc}\ N^a\ \pi^{bIJ}-\frac{N}{\sqrt{|\det
q|}}\ \eps_{abc}\ \pi^{aIK}\ \pi^{bJL}\ \eta_{KL}
\ee
with the solution $\pi^a_{IJ}=\eps^{abc}e_b^Ke_c^L\eps_{IJKL}$. First of
all, we check the first term
\be
-\frac{1}{2}\eps_{abc}\ N^a\ \pi^{b}_{IJ}
=\mp\frac{1}{2}\eps_{abc}N^a\eps^{bde}e_d^Ke_e^L\eps_{IJKL}=\pm\delta^d_a\delta^e_cN^ae_d^Ke_e^L\eps_{IJKL}=\pm
N^ae_a^Ke_c^L\eps_{IJKL}
\ee
And then the second term:
\be
&&\frac{N}{4\sqrt{|\det q|}}\ \eps_{abc}\ \pi^{a}_{IK}\ \pi^{b}_{JL}\
\eta^{KL}\nonumber\\
&=&\frac{N}{4\sqrt{|\det q|}}\eps_{abc}\eps^{ade}e_d^Me_e^N\eps_{IKMN}\
\eps^{bfh}e_f^Pe_h^Q\eps_{JLPQ}\ \eta^{KL}\ =\ -\frac{N}{2\sqrt{|\det
q|}}\delta_a^f\delta_c^h\eps^{ade}e_d^Me_e^Ne_f^Pe_h^Q\eps_{IKMN}\eps_{JLPQ}\
\eta^{KL}\nonumber\\
&=&-\frac{N}{2\sqrt{|\det
q|}}\eps^{dea}e_d^Me_e^Ne_a^Pe_c^Q\eps_{IKMN}\eps_{JLPQ}\
\eta^{KL}\nonumber
\ee
here now we define a new field $e^t_H$ or $n_H$ such that
$\eps^{dea}e_d^Me_e^Ne_a^P=\pm N\sqrt{|\det q|}\eps^{HMNP}e^t_H=\mp
\sqrt{|\det q|}\eps^{HMNP}n_H$ and $n^Hn_H=-1$. Thus
\be
&&\frac{N}{4\sqrt{|\det q|}}\ \eps_{abc}\ \pi^{a}_{IK}\ \pi^{b}_{JL}\
\eta^{KL}\nonumber\\
&=&\pm\frac{N}{2\sqrt{|\det q|}} \sqrt{|\det
q|}\eps^{HPMN}n_He_c^Q\eps_{IKMN}\eps_{JLPQ}\ \eta^{KL}\ =\ \mp
N\left[\delta^H_I\delta^P_K-\delta^H_K\delta^P_I\right]n_He_c^Q\eps_{JLPQ}\
\eta^{KL}\nonumber\\
&=&\pm Nn_Ke_c^Q\eps_{JLIQ}\ \eta^{KL}\ =\ \pm Nn^Le_c^Q\eps_{IJLQ}
\ee
As a result:
\be
X_{tc}^{IJ}=\pm\left(Nn^K+N^ae_a^K\right)e_c^L\eps^{IJ}_{\ \
KL}\equiv\pm e_t^Ke_c^L\eps^{IJ}_{\ \ KL}
\ee
To check the non-degeneracy of $e_\a^I$, we calculate its determinate
\be
\det e_\a^I&=&\frac{1}{4!}\eps^{\a\b\g\delta}e^I_\a e^J_\b e^K_\g
e^L_\delta\eps_{IJKL}\ =\ \frac{1}{3!}\eps^{tabc}e^I_t e^J_a e^K_b
e^L_c\eps_{IJKL}\ =\ \frac{1}{3!}\eps^{tabc}\left(Nn^I+N^de_d^I\right)
e^J_a e^K_b e^L_c\eps_{IJKL}\nonumber\\
&=&N\frac{1}{3!}\eps^{tabc}n^Ie^J_a e^K_b e^L_c\eps_{IJKL}\ =\ \pm
N\sqrt{|\det q|}
\ee
which is nonzero for non-vanished $N$. \finishproof}

By this Lemma, we can immediately simplify the expression of
Eq.(\ref{deltas}) to be
\be
&&\cv^9\delta^{3}\left(\frac{1}{2}N^a\left[\eps_{ab[c} \pi^{b0i}
X_{td]}^{jk}+\eps_{ab[c} \pi^{bij}X_{td]}^{0k}
\right]\eps_{ijk}-\frac{N}{4\sqrt{\det q}}\left[\eps_{ab[c} \pi^{a0l}
\pi^{bil} X_{td]}^{jk}+\eps_{ab[c} \pi^{aiK} \pi^{bjL}
\eta_{KL}X_{td]}^{0k}\right]\eps_{ijk}\right)\nonumber\\
&&\times \delta^{6}\left(X_{tc}^{IJ}\ X_{td}^{KL}\ \eps_{IJKL}\right)\
\delta^9\left(X_{tc}^{IJ}\
\pi^{dKL}\eps_{IJKL}-\frac{1}{3}\delta^d_cN\frac{1}{3!}\eps^{tabc}n^Ie^J_a
e^K_b e^L_c\eps_{IJKL}\right)
\ee
Now we are ready to integral over $N$ and $N^a$ and obtain
\be
\cz_\pm&=&\int_{II\pm}\cd\o_\a^{IJ}\cd X_{\a\b}^{IJ}\ \prod_{x\in M}\
\cv^{12} V_s\ \frac{1}{\left|\det \left(M\right)\right|}\
\delta^{20}\left(\eps_{IJKL}\ X_{\a\b}^{IJ}\
X_{\g\delta}^{KL}-\frac{1}{4!}\cv\eps_{\a\b\g\delta}\right)\sqrt{|D_1|}\prod_{\a}\delta(\xi_\a)\nonumber\\
&&\times\exp
i \int (\plform - \frac{1}{\gamma} \dual \plform)_{IJ} \wedge F^{IJ}
\ee
where
\be
M^e_a&:=&\left[\eps_{abc} \pi^{b0i} X_{td}^{jk}+\eps_{abc}
\pi^{bij}X_{td}^{0k} \right]\eps_{ijk}\eps^{cde}\nonumber\\
&=&\left[(\delta_a^d\delta_b^e-\delta_b^d\delta_a^e) \pi^{b0i}
X_{td}^{jk}+(\delta_a^d\delta_b^e-\delta_b^d\delta_a^e)\pi^{bij}X_{td}^{0k}
\right]\eps_{ijk}\nonumber\\
&=&\left[\pi^{e0i} X_{ta}^{jk}-\delta_a^e \pi^{b0i}
X_{tb}^{jk}+\pi^{eij}X_{ta}^{0k} -\delta_a^e\pi^{bij}X_{tb}^{0k}
\right]\eps_{ijk}\nonumber\\
&=&\frac{1}{6}\cv\delta^e_a-\frac{1}{2}\cv\delta^e_a=-\frac{1}{3}\cv\delta^e_a
\ee
Thus $|\det(M)|=\cv^3$ and the final result is
\be
\label{delta_final}
\cz_\pm&=&\int_{II\pm}\cd\o_\a^{IJ}\cd X_{\a\b}^{IJ}\ \prod_{x\in M}\
\cv^{9} V_s\ \delta^{20}\left(\eps_{IJKL}\ X_{\a\b}^{IJ}\
X_{\g\delta}^{KL}-\frac{1}{4!}\cv\eps_{\a\b\g\delta}\right)\sqrt{|D_1|}\prod_{\a}\delta(\xi_\a)\nonumber\\
&&\times\exp i \int (\plform - \frac{1}{\gamma} \dual \plform)_{IJ} \wedge F^{IJ}
\ee
which is resulting path-integral of Plebanski-Holst action on both
sector $(II\pm)$. Considering both sectors is the preparation for the
spin-foam construction.

\section{Consistency with Buffenoir, Henneaux, Noui and Roche}
\label{consisBHNR}

On setting $\gamma = \infty$, the path integral of the last section
becomes a Plebanski path integral.
However, at first glance, this path integral is different from the one
derived in the paper of
Buffenoir, Henneaux, Noui and Roche
(BHNR) \cite{bhnr}, having a different measure factor.
In this section we will show that this discrepancy is only apparent,
and show how the two path integrals are in fact equivalent.
Because in this section we set $\gamma = \infty$,
$\plconj = \plform$.

The key difference in the analysis of \cite{bhnr}
is that $B_{ta}$ is made into a \textit{dynamical variable} by
introducing a conjugate variable $P^\mu_{IJ}$ constrained,
however, to be zero.  This leads, in a precise sense,
to the presence of ``two lapses and two shifts.''
First, lapse and shift appear as certain components of
$B_{ta}$; we shall call these `physical lapse and shift' and
shall denote them $N_p, N_p^a$.
More precisely, we define $N_p, N_p^a$ to be functions of
$B_{ta}$ in the same way $N, N^a$ depended on $\plform_{ta}$ in the last
section.
A second lapse and shift appear on writing the
Hamiltonian and vector constraint delta functions in exponential form;
we shall call these
`langrange multiplier lapse and shift' and shall denote them $N, N^a$.
($N_p, N_p^a, N, N^a$ will always be undensitized.)

In the last section, by contrast, only one lapse and shift
appeared.  For this reason, the path integral
of \cite{bhnr} is not directly comparable with the path integral of the
last
section, but rather first one of the extra lapse and shift needs to
be removed before comparison.
We will carry this out in the present section, and see
how the path integral of \cite{bhnr} in fact reduces to that
of the last section.
Of course, we knew that these
two path integrals must be equivalent since they are constructed from a
single reduced phase space --- that of GR.
It is nevertheless instructive to see explicitly how the equivalence
comes about.
This also provides a valuable check against errors, by deriving the
final path
integral from two independent starting points.

\vspace{0.25cm}
\noindent\textit{Choice of gauge-fixings and manipulation of
constraints}

We start from equation (75) in \cite{bhnr}:
\begin{equation}
\label{frombhnr}
\scrZ = \int \scrD \omega_a^{IJ} \scrD \pi^a_{IJ} \scrD \beta_{a}^{IJ}
\left(\sqrt{D_1^{(\psi, \xi)} |_{P=0}} \delta(C_0') \delta(C_a')
\delta(G)\delta(\xi_\alpha)\right)
\left(N_p^{11} V_s^{15} \delta(C_{\pi\pi}) \delta(C_{\beta\pi})
\delta(C_{\beta\beta})
\delta(\tilde{T}) \right)
\exp i \int \rmd t \rmd^3 x \pi_{IJ}^a\dot{\conn}_a^{IJ},
\end{equation}
where we have used $\scrV = N_p V_s$ and $(\sqrt{D_2})_{BHNR} = N_p^{20}
V_s^{24}$, and
where we have also used the presence of constraint delta functions
to remove a weakly vanishing term $\tilde{H}$ that is present in the
exponent in
equation (75) of \cite{bhnr}. (Indeed, in \cite{bhnr},
$\tilde{H}$ is \textit{introduced} into the exponent in this way, using
that
$\tilde{H}$ vanishes weakly.)
Here $C_0'$, $C_a'$, and $\tilde{T}^{ab}$ are as defined in \cite{bhnr},
and are essentially
the scalar constraint, vector constraint, and secondary constraint
generated by $C_{\pi\pi}$, respectively.
$\xi_\alpha = (\xi_{\kappa_0}, \xi_{\kappa_a}, \xi_S, \xi_V, \xi_G)$
are the gauge-fixing functions corresponding to the full set of
first class constraints $\kappa_0, \kappa_a, C_0', C_a', G_{IJ}$
originally present in \cite{bhnr}.
(By the time one reaches the above equation in \cite{bhnr},
$\delta(\kappa_0)\delta(\kappa_a)$ have already been integrated out, but
their
corresponding gauge-fixing functions have not.)
To review, $\kappa_0$ and $\kappa_a$ are defined by
\begin{eqnarray}
\kappa_0 &:=& \half P^a_{IJ} B_{ta}^{IJ}  \\
\kappa_a &:=& \half \epsilon_{abc} P^b_{IJ} \pi^{cIJ}
\end{eqnarray}
and the gauge they generate correspond precisely to the freedom in the
choice
of physical lapse and shift $N_p, N_p^a$.
$D_1^{(\psi,\xi)}$ in the above path integral denotes the determinant of
the poisson bracket
matrix
\begin{displaymath}
\left(\begin{array}{cc}
\{\psi_\alpha, \psi_\beta\} & \{\psi_\alpha, \xi_\beta\} \\
\{\xi_\alpha, \psi_\beta\} & \{\xi_\alpha, \xi_\beta\}
\end{array}\right)
\end{displaymath}
where $\psi_\alpha$ collectively denotes the first class constraints
$\kappa_0, \kappa_a, C'_0, C'_a, G$, so that
$\{\psi_\alpha, \psi_\beta\} \approx 0$, and we have
$D_1^{(\psi, \xi)} = (\det \{\psi_\alpha, \xi_\beta\})^2$.
Because of the argument in
\cite{companion} without loss of generality we may assume
for convenience
a particular choice of gauge fixing:
\begin{eqnarray}
\xi_{\kappa_0} &:=& N_p - 1 \approx 0 \\
\xi_{\kappa_a} &:=& N_p^a \approx 0 .
\end{eqnarray}
With this choice, one can check
\begin{eqnarray*}
\{\kappa_0, \xi_{\kappa_0}\} = -\half N_p &\qquad& \{\kappa_0,
\xi_{\kappa_b}\} = -\half N_p^b  \\
\{\kappa_a, \xi_{\kappa_0}\} =  0 &\qquad& \{\kappa_a, \xi_{\kappa_b}\}
= -\delta^b_a .
\end{eqnarray*}
We furthermore assume that the gauge-fixing functions
$\xi_{S}, \xi_{V}, \xi_{G}$ are chosen to depend only
on $\pi^a_{IJ}$; this is clearly possible due to the fact
that the scalar, vector, and Gauss constraints are also
present in the Hamiltonian framework of Barros e Sa \cite{barros}
derived from the Holst action, and there it is possible
to choose pure momentum gauge-fixing conditions,
hence depending only on $\pi^a_{IJ}$.

Second, recall that
\begin{equation}
\sqrt{D_1^{(\psi, \xi)}} \prod_{\alpha} \delta(\psi_\alpha)
\delta(\xi_\alpha)
\end{equation}
is invariant under the choice of functions $\psi_\alpha, \xi_\alpha$
enforcing the
chosen gauge-fixed constraint surface.
We use this to replace $C_0', C_a'$ in favor of the constraints
$H, H_a$ defined in the
foregoing sections.
That this replacement is valid can be seen
in two steps:
\begin{enumerate}
\item Replace $C_0', C_a'$ with $C_0, C_a$ as defined
in \cite{bhnr}.
These differ from $C_0', C_a'$ by a linear combination
of the other constraints (see \cite{bhnr}).
\item
Within the constraint surface defined by the simplicity constraints and
$C_a \approx 0$, we have $C_0 = N_p H$.
This, combined with $N_p \neq 0$\footnote{BHNR
\cite{bhnr} assumes non-degeneracy
of the 4-metric, which implies $N_p \neq 0$.
Of course there is some hand-waving here, because in fact BHNR
integrates over all possible $N_p$ in the path integral.
},
allows one to replace $C_0$ by $H$. Lastly, $C_a$ is equal to $H_a$.
\end{enumerate}
Let $\tilde{\psi}_\alpha$ denote the new constraint functions
$\kappa_0, \kappa_a, H, H_a, G_{IJ}$.

The assumptions about the gauge-fixing conditions
imply that the
poisson-bracket matrix $\{\tilde{\psi}_\alpha, \xi_\beta\}$
is of the form
\begin{equation}
\begin{array}{rccccc}
&\xi_{\kappa_0} & \xi_{\kappa_b} & \xi_S & \xi_V & \xi_G \\
\cline{2-6}
\multicolumn{1}{r|}{\kappa_0} &
\multicolumn{1}{c|}{- \half N_p} &
\multicolumn{1}{c|}{- \half N_p^b} &
& \multirow{2}{*}{0} & \multicolumn{1}{c|}{} \\
\cline{2-3}
\multicolumn{1}{r|}{\kappa_a} &
\multicolumn{1}{c|}{0} &
\multicolumn{1}{c|}{-\delta^b_a} &
\multicolumn{3}{c|}{} \\
\cline{2-6}
\multicolumn{1}{r|}{H} &
\multicolumn{2}{c|}{} &
\multicolumn{3}{c|}{} \\
\multicolumn{1}{r|}{H_a} &
\multicolumn{2}{c|}{B} & & A &
\multicolumn{1}{c|}{} \\
\multicolumn{1}{r|}{G} &
\multicolumn{2}{c|}{} &
\multicolumn{3}{c|}{} \\
\cline{2-6}
\end{array}.
\end{equation}
The fact that we are now using $H, H_a$
ensures that $A$ is independent of $N_p, N_p^a$.  Thus,
we have the factorization
\begin{equation}
D_1^{(\tilde{\psi}, \xi)} = \left(\det \{\psi_\alpha,
\xi_\beta\}\right)^2 = \frac{1}{4} (N_p \det A)^2
\end{equation}
with $\det A$ independent of $N_p, N_p^a$.
In fact, if we choose the gauge-fixings in the Holst
path integral to be the same as the gauge-fixings $\xi_S, \xi_V, \xi_G$
then $\det A = \sqrt{D_1^{Holst}}$.  We will write $\sqrt{D_1^{Holst}}$
from now on.  This gives us
\begin{equation}
Z= \int \scrD \omega_a^{IJ} \scrD \pi^a_{IJ} \scrD \beta_a^{IJ}
\left(N_p \sqrt{D_1^{Holst}} \delta(H) \delta(H_a) \delta(G)
\delta(\xi_\alpha) \right)
\left(N_p^{11} V_s^{15} \delta(C_{\pi\pi}) \delta(C_{\beta\pi})
\delta(C_{\beta\beta})
\delta(\tilde{T}) \right)
\exp i \int \rmd t \rmd^3 x \left[ \pi^a_{IJ} \dot{\omega}^{IJ}_a
\right]
\end{equation}
Finally, when the other constraints are satisfied,
$\tilde{T}^{ab} = N_p D^{ab}$, where $D^{ab}$ is as in
(\ref{holstconstr}).
Thus we may replace $\delta(\tilde{T})$ by $\delta(N_p D^{ab}) =
\frac{1}{N_p} \delta(D^{ab})$,
\begin{equation}
Z= \int \scrD \omega_a^{IJ} \scrD \pi^a_{IJ} \scrD \beta_a^{IJ}
\left(N_p \sqrt{D_1^{Holst}} \delta(H) \delta(H_a) \delta(G)
\delta(\xi_\alpha) \right)
\left(N_p^{10} V_s^{15} \delta(C_{\pi\pi}) \delta(C_{\beta\pi})
\delta(C_{\beta\beta})
\delta(D^{ab}) \right)
\exp i \int \rmd t \rmd^3 x \left[ \pi^a_{IJ} \dot{\omega}^{IJ}_a
\right] .
\end{equation}

\vspace{0.25cm}
\noindent\textit{Integrating out $N_p, N_p^a$:}

From (\ref{final_dBIJ})
\begin{displaymath}
\scrD \beta_a^{IJ} = \scrV^{-6} V_s
\scrD N_p \scrD N_p^a \scrD \tilde{C}_{\beta\pi} \scrD
\tilde{C}_{\beta\beta} ,
\end{displaymath}
so that we have
\begin{equation}
\scrZ = \int \scrD \omega_a^{IJ}
\scrD \pi^a_{IJ} \scrD N_p \scrD N_p^a \scrD \tilde{C}_{\beta\pi} \scrD
\tilde{C}_{\beta\beta}
N_p^{5} V_s^{10}
\left(\sqrt{D_1^{Holst}} \delta(H) \delta(H_a) \delta(G)
\delta(\xi_\alpha) \right) \delta(C_{\pi\pi})
\delta(C_{\beta\pi}) \delta(C_{\beta\beta}) \delta(D^{ab})
\exp i \int \rmd t \rmd^3 x
\left[\pi_{IJ}^a\dot{\conn}_a^{IJ} \right].
\end{equation}
The only $N_p, N_p^a$ dependence is that explicitly shown above.
Factoring out the the integrals over $N_p, N_p^a$ and evaluating gives
\begin{equation}
\int \scrD N_p \scrD N_p^a N_p^{5}
\delta(\xi_{\kappa_0})\delta(\xi_{\kappa_a}) = 1.
\end{equation}
Inserting this gives
\begin{equation}
Z= \int \scrD \omega_a^{IJ} \scrD \pi^a_{IJ} \scrD \tilde{C}_{\beta\pi}
\scrD \tilde{C}_{\beta\beta} V_s^{10}
\left(\sqrt{D_1^{Holst}} \delta(H) \delta(H_a) \delta(G) \delta(\xi_S)
\delta(\xi_V) \delta(\xi_G) \right)
\delta(C_{\pi\pi}) \delta(C_{\beta\pi}) \delta(C_{\beta\beta})
\delta(D^{ab})
\exp i \int \rmd t \rmd^3 x \left[ \pi^a_{IJ} \dot{\omega}^{IJ}_a
\right] .
\end{equation}
Using the inverse of the reasoning leading to (\ref{plebpath_mid}), we replace
$C_{\beta\pi}$ with $\tilde{C}_{\beta\pi}$, then $C_{\beta\beta}$ with
$\tilde{C}_{\beta\beta}$,
and integrate out $\tilde{C}_{\beta\pi}$, $\tilde{C}_{\beta\beta}$.
This yields
\begin{equation}
Z= \int \scrD \omega_a^{IJ} \scrD \pi^a_{IJ} V_s^{10}
\left(\sqrt{D_1^{Holst}} \delta(H) \delta(H_a) \delta(G) \delta(\xi_S)
\delta(\xi_V) \delta(\xi_G) \right)
\delta(C_{\pi\pi}) \delta(D^{ab})
\exp i \int \rmd t \rmd^3 x \left[ \pi^a_{IJ} \dot{\omega}^{IJ}_a
\right] .
\end{equation}
which is precisely equation (\ref{holstphase}), which in the last
section
was in turn shown to be equal to (\ref{plebpath_final}).

\vspace{0.25cm}
\noindent\textit{Remark on BHNR}

In the above we began with equation (75) in BHNR \cite{bhnr}, and not
the final answer (76) in BHNR.  This is because in
BHNR the Henneaux-Slavnov trick \cite{hs} was not applied correctly
in passing from (75) to (76).
Specifically, as already mentioned earlier in this section, BHNR
introduces the Hamiltonian constraint into the path integral exponential
`for free' by using the presence of $\delta(C_0)$ in the path integral,
instead of by `exponentiating' $\delta(C_0)$.  As a consequence,
$\delta(C_0)$ remains explicitly in the path integral; but $\delta(C_0)$
is not invariant under the canonical transformation
used in the Henneaux-Slavnov trick introduced in \cite{bhnr}, even
on-shell. (\cite{bhnr} explicitly calculates the change of $C_0$ under
the canonical transformation.) This was overlooked in \cite{bhnr} and
presents an obstacle to using the Henneaux-Slavnov trick.

In section \ref{holst_phase} of the present paper, however,
the Hamiltonian constraint is brought into the exponential
by casting the associated delta function in exponential form.
As a consequence, no similar problem arises when performing
the Henneaux-Slavnov trick, and the trick goes through.

\section{Discussion}

The goal of the present work has been to calculate the appropriate
formal path integrals for Holst gravity and for Plebanski gravity with
Immirzi parameter
--- which we call Holst-Plebanski gravity --- as determined by canonical
analysis. This has been done, starting from the $SO(\eta)$ covariant
framework of \cite{barros}.
The final Holst-Plebanski path integral was shown to
be consistent with the calculations of \cite{bhnr}, modulo a slight
oversight in \cite{bhnr} which we corrected. We used the well known
reduced phase space method \cite{HT} in our derivation of which a
compact account adapted to the notation employed here can be found in
\cite{companion}.

The main difference between the formal path integral expression for
Holst gravity derived in this paper and the ``new spin foam models''
\cite{spinfoam2} that are also supposed to be quantisations of Holst
gravity are\footnote{The origin of the differences is that the approach in the present paper relies on the canonical quantization, while the standard spin-foam approach doesn't.} 1. the appearence of the local measure factor, 2. the
continuum rather than discrete formulation (triangulation) and the lack
of manifest spacetime covariance\footnote{In fact discrete models are
also never
spacetime covariant unless they are topological, however, the
continuum limit of spin foam models, if one could take it, should be
spacetime covariant.}. The next steps in
our programme are therefore clear: In \cite{muxin2} we propose a
discretisation of the path integral derived in this paper which does
take the proper measure factor into account. We will do this using a
new method designed to take care of the simplicity constraints of
Plebanski
gravity and which lies somewhere between the spirits of
\cite{spinfoam2} and \cite{BFT}. As we have said before, we interpret
the lack of manifest spacetime covariance even in the continuum
as an unavoidable consequence of the mixture of dynamics and gauge
invariance in background independent (generally covariant) theories
with propagating degrees of freedom.
In the classical theory it requires some work to establish that
spacetime covariance actually does hold on, albeit on shell only.
However, the quantum
corrections apparently depend on the off shell physics and thus lack of
spacetime diffeomorphism invariance may well prevail outside
the semiclassical regime. 
A question
is whether there is a different symmetry group of the quantum
theory, which coincides with spacetime diffeomorphism invariance on
shell in the classical theory. The obvious candidate for this
``quantum diffeomorphism group'' is the quantisation of the Bergmann --
Komar group (BKG) \cite{BK} as was proposed in \cite{thiemannbook} and
in \cite{muxin4} it is analysed if and in
which sense the BKG is a symmetry of the quantum theory.

\section*{Acknowledgments}

J.E. thanks Eugenio Bianchi for discussions, and gratefully acknowledges
partial support
by NSF grant OISE-0601844 and the Alexander von Humboldt foundation of
Germany. M.H. thanks Aristide Baratin, Michael K\"ohn, Florian L\"obbert
and Yongge Ma for discussions, and acknowledges the support by
International Max Planck Research School and the partial support by NSFC Nos. 10675019 and 10975017.

\appendix

\section{An example for checking the equivalences between the
path-integrals of the Holst Hamiltonian, Ashtekar-Barbero-Immirzi
Hamiltonian and triad-ADM Hamiltonian formalisms: Imposing the
time-gauge}\label{check}

In order to check these equivalences we need to fix the boost part of
the internal gauge transformations by imposing the time-gauge, i.e. inserting
the delta function $\delta(e_a^0)$ and the corresponding Faddeev-Popov
determinant into the path-integral formula Eq.(\ref{pi}). In order
to do that, we need the time gauge condition written in terms of
canonical variables of Holst action. But it is not hard to find:
\be
-2(\det
e_a^i)e_c^0=\tilde{\pi}^{ai}\pi^b_i\eps_{bac}
=-\frac{1}{8}\eps^{ijk}\pi^a_{0i}\pi^{b}_{jk}\eps_{abc}
\ee
then we denote the time-gauge condition
$T_c:=\eps^{ijk}\pi^a_{0i}\pi^{b}_{jk}\eps_{abc}=0$ instead of
$e_c^0=0$, in the sector that $e_a^i$ is non degenerate. On the other
hand, the boost part of the Gauss constraint reads:
\be
G_{0j}&=&\partial_a(\pi-\frac{1}{\g}*\pi)^a_{0j}+\o_{a0}^{\ \
k}\pi^a_{jk}-\o_{aj}^{\ \ k}\pi^a_{0k}
\ee
The Faddeev-Popov determinant $\Delta_{FP}$ is defined by:
\be
\Delta_{FP}=\left|\det\left(\{G_{0i}(x), T_c(x')\}_D\right)\right|
\ee
where $\{\ ,\ \}_D$ is the Dirac bracket with respect to the second
class constraints $C^{ab}$ and $D^{ab}$. However since both $C^{ab}$ and
$D^{ab}$ are $SO(\eta)$ gauge invariant, they are Poisson-commutative
with the Gauss constraint $G_{IJ}$. Therefore the Dirac bracket between
$G_{0i}$ and $T_c$ is identical to their Poisson bracket, so that
\be
\{G_{0j}(x), T_c(x')\}_D&=&\{G_{0j}(x), T_b(x')\}\ =\ \left\{
\left(\o_{a0}^{\ \ k}\pi^a_{jk}-\o_{aj}^{\ \ k}\pi^a_{0k}\right)(x),\
\eps^{imn}\pi^d_{0i}\pi^{b}_{mn}\eps_{dbc}(x')\right\}\nonumber\\
&=&-\eps_{dbc}\eps^{imn}\pi^a_{jk}\pi^{b}_{mn}\left\{ \o_{a}^{0k}(x),\
\pi^d_{0i}(x')\right\}-\eps_{dbc}\eps^{imn}\pi^a_{0k}\pi^d_{0i}\left\{
\o_{a}^{jk}(x),\ \pi^{b}_{mn}(x')\right\}\nonumber\\
&=&-\eps_{dbc}\eps^{imn}\pi^a_{jk}\pi^{b}_{mn}\delta^d_a\left(\delta^0_0\delta^k_i-\delta^0_i\delta^k_0\right)\delta(x,x')-\eps_{dbc}\eps^{imn}\pi^a_{0k}\pi^d_{0i}\delta^b_a\left(\delta^j_m\delta^k_n-\delta^j_n\delta^k_m\right)\delta(x,x')\nonumber\\
&=&\eps_{abc}\eps^{imn}\pi^a_{ij}\pi^{b}_{mn}\delta(x,x')-2\eps_{abc}\eps^{ijk}\pi^a_{0i}\pi^b_{0k}\delta(x,x')\nonumber
\ee
where the first term vanishes by the time-gauge $e_a^0=0$ (our analysis is
for the sector of solutions in which
$\pi^a_{IJ}=\eps^{abc}e_b^Ke_c^L\eps_{IJKL}$). And the second term
\be
-2\eps_{abc}\eps^{ijk}\pi^a_{0i}\pi^b_{0k}=-2\left[\det
e_a^i\right]^2\eps_{abc}\eps^{ijk}f^a_if^b_k=-2\left[\det
e_a^i\right]\eps_{ikl}\eps^{ijk}e^l_c=4\left[\det e_a^i\right]e_c^j
\ee
As a result, we obtain that $\Delta_{FP}=\prod_{x\in M}\left[\det
e_a^i\right]^4$. Thus we can insert the gauge fixing term
$\Delta_{FP}\delta^3(T_c)$ into the phase space path-integral
Eq.(\ref{pi}).
\be
Z_T&=&\int[D\o_a^{IJ}][D\pi_{IJ}^a]\ \prod_{x\in M}\ \delta(G^{IJ})\
\delta(H_a)\ \delta(H)\ \delta(C^{ab})\ \delta(D^{ab})\ \delta(T_c)\
\Delta_{FP}\ \sqrt{|D_2|}\ \exp i\int\rmd t\int\rmd^3x\
\pi_{IJ}^a\dot{\o}_a^{IJ}\nonumber\\
&=&\int[D\o_a^{IJ}][D\pi_{IJ}^a]\ \prod_{x\in M}\ \delta(G^{IJ})\
\delta(H_a)\ \delta(H)\ \delta(C^{ab})\ \delta(D^{ab})\ \delta(T_c)\
V_s^{14}\ \exp i\int\rmd t\int\rmd^3x\
\pi_{IJ}^a\dot{\o}_a^{IJ}\label{TGCPI}
\ee
Following the same strategy as in section \ref{config_holst}, we can obtain the
time-gauge-fixed path-integral for the Holst action which is expressed
as
\be
Z_T&=&\int[D\conn_\a^{IJ}][De^I_\a]\ \prod_{x\in M}\ \cv^3V_s^5\
\delta^3\left((\det e_a^i)e_c^0\right)\ \exp i\int\ e^I\wedge
e^J\wedge\left(*F_{IJ}-\frac{1}{\gamma}F_{IJ}\right)[\conn]\nonumber\\
&=&\int[D\conn_\a^{IJ}][De^I_\a]\ \prod_{x\in M}\ \cv^3V_s^2\
\delta^3\left(e_c^0\right)\ \exp i\int\ e^I\wedge
e^J\wedge\left(*F_{IJ}-\frac{1}{\gamma}F_{IJ}\right)[\conn]\label{pi7}
\ee
where $\cv=|\det e_\a^I|$ is the 4-dimensional volume element, and
$V_s=|\det e_a^i|$ is the spatial volume element when we have imposed
the time-gauge fixing condition.

In the following, starting from Eq.(\ref{TGCPI}), we try to derive a canonical path-integral formula for
the Hamiltonian framework used in canonical LQG, i.e. the
Ashtekar-Barbero-Immirzi Hamiltonian. The product of two $\delta$-functions
in (\ref{TGCPI}) can be rewritten
\be
\delta\left(T_c\right)\delta\left(C^{ab}\right)&=&\delta^3\left(\eps^{ijk}\pi^a_{0i}\pi^{b}_{jk}\eps_{abc}\right)\delta^6\left(\eps^{ijk}\pi^{(a}_{0i}\pi^{b)}_{jk}\right)\
=\ \delta^9\left(\eps^{ijk}\pi^a_{0i}\pi^{b}_{jk}\right)\nonumber\\
&=&\left[\det
\pi^a_{0i}\right]^{-3}\delta^{9}\left(\pi^{b}_{jk}\right)=V_s^{-6}\delta^{9}\left(\pi^{b}_{jk}\right)
\ee
then we integrate over $\pi^{b}_{jk}$ in Eq.(\ref{TGCPI}) and denote
$\pi^a_{0i}=\sqrt{|\det q|}f^a_i\equiv E^a_i$
\be
Z_T&=&\int[DA_a^i][D\G_a^i][DE_i^a]\ \prod_{x\in M}\ \delta(G^{IJ})\
\delta(H_a)\ \delta(H)\ \delta(D^{ab})\ V_s^{8}\ \exp i\int\rmd
t\int\rmd^3x\ E_i^a\dot{A}_a^{i}
\ee
where $A^i_a\equiv\o^{0i}=A_a^{0i}-\frac{1}{2\gamma}\eps^{ijk}A_a^{jk}$
and $\G^i_a\equiv\frac{1}{2}\eps^{ijk}A_a^{jk}$. We then obtain the
relation:
\be
\o_a^{ij}=\eps^{ijk}\left[(1+\g^{-2})\G^k_a+\g^{-1}A^k_a\right]
\ee
Then the Gauss constraint $G_{IJ}$ and the secondary constraint $D^{ab}$
become
\be
G_{0i}&=&\partial_aE^a_i+\eps^{ijk}\left[(1+\g^{-2})\G^j_a+\g^{-1}A^j_a\right]E_k^a\nonumber\\
G_i&=&\frac{1}{2}\eps_{ijk}G_{ij}\ =\
-\g^{-1}\partial_aE^a_k+\eps_{kij}A_{a}^iE^a_{j}\nonumber\\
D^{ab}&=&\frac{1}{\sqrt{|\det
q|}}\left[\eps_{ijk}E^c_i\left(\partial_cE^a_j\right)E^b_k+\eps_{ijk}E^c_i\left(\partial_cE^b_j\right)E^a_k+\eps_{ijk}E^c_i\eps_{jmn}\G^m_cE^a_nE^b_k+\eps_{ijk}E^c_i\eps_{jmn}\G^m_cE^b_nE^a_k\right]\nonumber\\
\ee
And the result in \cite{barros} means that
\be
\delta^{6}\left(G^{IJ}\right)\delta^6\left(D^{ab}\right)=\left[\frac{\partial\left(G_{0i},D^{ab}\right)}{\partial
\G_a^i}\right]^{-1}\delta^9\left(\G_{a}^i-\frac{1}{2}\eps_{ijk}E^{b}_j\left[\partial_{[b}E_{a]}^k+E_{a[l}E^c_{k]}\partial_bE_{cl}\right]\right)\delta^3\left(G_i\right)
\ee
which shows that $\G^i_a$ is the spin-connection compatible with the
triad $E^a_i$, and we have defined $E_a^i=e_a^i/\sqrt{|\det q|}$ as the
inverse of $E^a_i$. In order to compute the Jacobian factor, we observe
that:
\be
&&\delta^6\left(D^{ab}\right)\delta^{3}\left(G_{0i}\right)
\ =\ {\left|\det\
E_{(a}^iE_{b)}^j\right|}\delta^6\left(D^{ab}E_a^iE_b^j\right)\delta^{3}\left(\frac{\g}{1+\g^2}\left[\g
G_{0i}-G_i\right]\right)\nonumber\\
&=&V_s^{-8}\delta^6\left(\frac{4}{\sqrt{|\det
q|}}\left[E^a_{(i}\G_{|a|j)}-\delta_{ij}E^a_k\G^k_a\right]+\cdots\right)\delta^{3}\left(\eps_{ijk}\G_{aj}E^{a}_k+\cdots\right)\nonumber\\
&=&V_s^{-2}\delta^6\left(4\left[E^a_{(i}\G_{|a|j)}-\delta_{ij}E^a_k\G^k_a\right]+\cdots\right)\delta^{3}\left(\eps_{ijk}\G_{aj}E^{a}_k+\cdots\right)\nonumber\\
&=&V_s^{-2}\delta\left(4\left[E^a_{k}\G_{a}^k-3E^a_k\G^k_a\right]+\cdots\right)\delta^5\left(4\left[E^a_{(i}\G_{|a|j)}-\frac{1}{3}\delta_{ij}E^a_k\G^k_a\right]+\cdots\right)\delta^{3}\left(\eps_{ijk}\G_{aj}E^{a}_k+\cdots\right)
\ =\ V_s^{-2}\delta^9\left(E^a_i\G_{aj}+\cdots\right)\nonumber\\
&=&\frac{V_s^{-2}}{|\det
E^a_i|^3}\delta^9\left(\G_{a}^i-\frac{1}{2}\eps_{ijk}E^{b}_j\left[\partial_{[b}E_{a]}^k+E_{a[l}E^c_{k]}\partial_bE_{cl}\right]\right)
\ =\
V_s^{-8}\delta^9\left(\G_{a}^i-\frac{1}{2}\eps_{ijk}E^{b}_j\left[\partial_{[b}E_{a]}^k+E_{a[l}E^c_{k]}\partial_bE_{cl}\right]\right)
\ee
where in the fourth step we split the six $\delta$-functions for a
symmetric matrix into a $\delta$-function of its trace product five
$\delta$-functions of its traceless part. With this result, we can
integral over $\G_a^i$ and obtain the desired result:
\be
Z_T&=&\int[DA_a^i][DE_i^a]\ \prod_{x\in M}\ \delta(G_i)\ \delta(H_a)\
\delta(H)\ \exp i\int\rmd t\int\rmd^3x\ E_i^a\dot{A}_a^{i}\label{}
\ee
where the constraint $G_i$, $H_a$ and $H$ take the form as we used in
canonical LQG $(s=\eta_{00})$
\be
G_i&=&\partial_aE^a_i+\eps_{ijk}A_a^jE^a_k\ \ \ \ \ \ \ \ \ \ \ \
H_a\ =\ F_{ab}^iE^b_i-A_a^iG_i\nonumber\\
H&=&\frac{1}{2}\frac{E^a_iE^b_j}{\sqrt{|\det q|}}\left[\eps^{ij}_{\ \
k}F^k_{ab}-2\left(\g^2-s\right)K_{[a}^iK_{b]}^j\right]
\ee
Therefore we obtain equivalence between the time-gauge fixed Holst
action path-integral Eq.(\ref{pi7}) and the canonical path-integral of
the Ashtekar-Barbero-Immirzi Hamiltonian. Furthermore, the
Ashtekar-Barbero-Immirzi Hamiltonian formalism is symplectically
equivalent
to the triad ADM Hamiltonian formalism, i.e. there is a canonical
transformation $A_a^i=\G_a^i+\g K_a^i$ that relates the conjugate pair
$(A_a^i,E^a_i)$ to $(E^a_i,K_a^i)$, where the $su(2)$-valued 1-form
$K_a^i$ relates to the extrinsic curvature via $-sK_{ab}=K_{(a}^ie_{b)}^i$.
Then it is trivial that the time-gauge fixed Holst action path-integral
Eq.(\ref{pi7}) is also equivalent to the canonical path-integral of the
triad ADM Hamiltonian, i.e.
\be
Z_T&=&\int[DK_a^i][DE_i^a]\ \prod_{x\in M}\ \delta(G_i)\ \delta(H_a)\
\delta(H)\ \exp i\int\rmd t\int\rmd^3x\ {K}_a^{i}\dot{E}_i^a
\ee
where the constraint $G_i$, $H_a$ and $H$ takes the form:
\be
G_i&=&\eps_{ijk}K_a^jE^a_k\ \ \ \ \ \ \ \ \ \ \ \
H_a\ =\ 2sD_b\left[K_a^jE^b_j-\delta^b_aK_c^jE^c_j\right]\nonumber\\
H&=&-\frac{s}{\sqrt{|\det
q|}}\left(K_a^iK_b^j-K_b^iK_a^j\right)E^a_iE^b_j-\sqrt{|\det q|}R
\ee

\end{document}